\documentclass[12pt,oneside, a4paper]{article}

\ifx\pdfoutput\undefined
\usepackage[dvips,bookmarks=false]{hyperref}	
\else
\usepackage{hyperref}	
\fi
\hypersetup{colorlinks,bookmarksopen,bookmarksnumbered,citecolor=blue,
linkcolor=black,pdfstartview=FitH,urlcolor=blue}

\usepackage{graphicx}
\usepackage{amssymb}
\usepackage{cite}
\usepackage{bm}
\usepackage{indentfirst}
\usepackage{amsmath}
\usepackage{hhline}
\usepackage{multirow}

\allowdisplaybreaks

\usepackage{ulem}

\begin{document}

\begin{titlepage}

\begin{flushright}
KUNS-2760
\end{flushright}

\begin{center}

\vspace{1cm}
{\large\bf 
A $\mu$-$\tau$-philic scalar doublet under $Z_{n}$ flavor symmetry
 }
\vspace{1cm}

\renewcommand{\thefootnote}{\fnsymbol{footnote}}
Yoshihiko Abe$^{1}$\footnote[1]{y.abe@gauge.scphys.kyoto-u.ac.jp}
,
Takashi Toma$^{1,2}$\footnote[2]{takashi.toma@physics.mcgill.ca}
,
Koji Tsumura$^{1}$\footnote[3]{ko2@gauge.scphys.kyoto-u.ac.jp}
\vspace{5mm}

{\it%
$^1${Department of Physics, Kyoto University, Kyoto 606-8502, Japan}\\
 $^2${Department of Physics, McGill University,\\
 3600 Rue University, Montr\'{e}al, Qu\'{e}bec H3A 2T8, Canada}
}

\vspace{8mm}

\abstract{
 We propose a minimal model which accommodates the long-standing anomaly of muon magnetic moment 
 based on abelian discrete flavor symmetries.  
 The standard model is extended by 
 scalar doublets charged under a $Z_n$ lepton flavor symmetry. 
 In these models,  a large contribution to the muon magnetic moment 
 can be obtained by the chirality enhancement 
 from new scalar mediated diagrams without conflicting with the flavor symmetry.  
 Thanks to the lepton flavor symmetry, these models automatically 
 forbid lepton flavor violation. 
 The minimal model is based on $Z_4$ symmetry with only one extra scalar doublet. 
 In this model, we show that the parameter space favored by the muon $g-2$ 
 can easily be consistent with experimental constraints and 
 theoretical bounds such as the electroweak precision tests, 
 lepton universality, potential stability condition and triviality bound 
 as well as the LHC direct search mass bound. 
 The new contributions to the muon electric dipole moment and 
 the Higgs decay into $\gamma\gamma$ can be indirect signals of the model. 
 }

\end{center}
\end{titlepage}

\renewcommand{\thefootnote}{\arabic{footnote}}
\newcommand{\bhline}[1]{\noalign{\hrule height #1}}
\newcommand{\bvline}[1]{\vrule width #1}

\setcounter{footnote}{0}

\setcounter{page}{1}

\section{Introduction}
The Standard Model (SM) has been established by the discovery of the
Higgs boson at the LHC. 
New particles beyond the SM are also being searched at the LHC. 
However, there is no signature of new particles until now, 
and the experimental results are consistent with the SM predictions. 
Other than the high energy frontier experiment, 
many of flavor observables are measured very precisely as the luminosity frontier experiment. 
A striking indication of the beyond the SM would be 
the muon anomalous magnetic moment (muon $g-2$). 
There is a discrepancy between the measured value and the SM
prediction as~\cite{Tanabashi:2018oca}
%
\begin{align}
 \Delta a_{\mu} =
  a_{\mu}^\mathrm{exp}-a_{\mu}^\mathrm{SM}=268(63)(43)\times10^{-11}, 
\end{align}
%
where the numbers in the first and second parentheses represent 
the statistical and systematic errors, respectively. 
The total significance of the deviation is $3.5\sigma$ far from the SM prediction.\footnote{
A new evaluation of the hadronic vacuum polarization with recent experimental data gives 
a 3.7$\sigma$ deviation from the SM~\cite{Keshavarzi:2018mgv}. 
We here use the averaged value  obtained by PDG~\cite{Tanabashi:2018oca}. }
Note that there is a non-negligible large theoretical uncertainties 
in the hadronic contribution due to the light-by-light scattering~\cite{Prades:2009tw}. 
Currently, FNAL E989 experiment is ongoing, and will achieve a factor four
improvement on its precision at the end of the running~\cite{Chapelain:2017syu}.

There are many attempts to explain the discrepancy of the muon magnetic moment. 
For instance, in lepton-specific two Higgs doublet models (THDMs), the
new contribution to the muon magnetic moment due to the additional Higgs
bosons can be enhanced by a large $\tan\beta$, 
which is the ratio of vacuum expectation values (VEVs) of two Higgs doublets~\cite{Abe:2015oca,
Chun:2016hzs, Abe:2017jqo, Crivellin:2019dun}. 
The THDM with tree-level flavor changing neutral currents has also been 
studied to explain the discrepancy in the light of the $h\to\mu\tau$
excess at the LHC~\cite{Omura:2015nja, Omura:2015xcg}, which has been
disappeared. 
Another way is to consider a light $Z^\prime$ gauge
boson associated with an extra $U(1)_{{\mathbf L}_{\mu}-{\mathbf L}_{\tau}}$ 
symmetry~\cite{Baek:2001kca, Ma:2001md}, or a light hidden
photon~\cite{Endo:2012hp}. 
In these models, thanks to the new light mediator running in the loop
diagram, the muon magnetic moment can be enhanced even with a smaller coupling strength. 
There is also argument to account for the discrepancy in framework of
supersymmetry~\cite{Endo:2013bba}, axion-like
particle~\cite{Marciano:2016yhf} and fourth generation of
leptons~\cite{BarShalom:2011bb}.

In this paper, we propose a minimal model explaining the discrepancy
of the muon anomalous magnetic moment between the SM prediction and the
measurement based on abelian discrete symmetries. 
Models based on the abelian discrete groups easily give a sufficiently large 
and the correct sign of the contribution to the muon magnetic moment. 
The model based on a $Z_{4}$ symmetry is identified as the minimal model, 
which is a kind of variant of the inert scalar doublet model based on a $Z_{2}$ symmetry. 
Thanks to the $Z_4$ symmetry, the lepton flavor violating (LFV) processes such as
$\ell\to\ell^\prime\gamma~(\ell,\ell^\prime=e,\mu,\tau)$ are forbidden automatically 
against severe bounds of their non-observation. 
As a result, we find a solution to the muon $g-2$ anomaly 
without conflicting with the constraints from the electroweak
precision tests and the lepton universality of heavy charged lepton
decays and $Z$ boson leptonic decays. 
In addition, we examine whether the model is consistent with theoretical
bounds of potential stability and triviality.
We will formulate analytic expressions of these quantities, and
numerically explore the parameter space which can accommodate 
the discrepancy of the muon anomalous magnetic moment. 
As further perspective, neutrino mass generation mechanism and 
a distinctive collider signature, a prediction for muon electric dipole
moment induced by new CP phases and influence on the Higgs decay into
$\gamma\gamma$ will also be discussed.

\section{Flavor Charged Scalar Doublets}
\label{sec:1.1}
\begin{figure}[t]
\centering
   \includegraphics[scale=1.1]{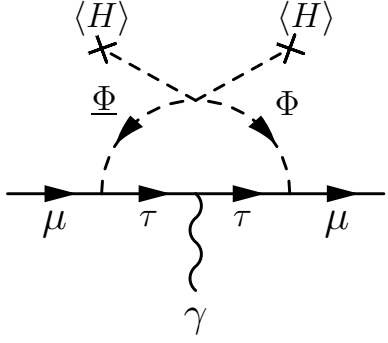}
   \caption{Feynman diagram inducing muon anomalous magnetic moment in
   $U(1)_{{\mathbf L}_{\mu}-{\mathbf L}_{\tau}}$ and $Z_n~(n=2,3,\cdots)$ models.}
   \label{fig:1}
 \end{figure}

Let us discuss a simple extension of the SM with a pair of scalar doublets 
$(\Phi, \underline{\Phi}\,)$ 
whose global $U(1)_{{\mathbf L}_{\mu}-{\mathbf L}_{\tau}}$ flavor charge is $(2,-2)$, 
where ${\mathbf L}_{\mu}$ and ${\mathbf L}_{\tau}$ represent the muon and tau lepton 
flavor numbers, respectively. 
Detailed quantum charge assignments are given in Table~\ref{Tab:Zn}. 
\begin{table}[tb]
\centering
\begin{tabular}{|c||c|c|c|c|c|c|}
\hline 
Particle & SM & $U(1)_{{\mathbf L}_{\mu}-{\mathbf L}_{\tau}}$ & $Z_{2}$ & $Z_{3}$ & $Z_{4}$ & $Z_{n}$ \\ 
\hline \hline
$(L_{e}, L_{\mu}, L_{\tau}) 
$ & $(1, 2)_{-1/2}$ & $(0, +1, -1)$ 
& $(+, -, -)$
& $(1, \omega, \omega^{2})$
& $(1, i, -i)$
& $(1, \omega, \underline{\omega})$ \\ 
$(e_{R}^{}, \mu_{R}^{}, \tau_{R}^{})$ & $(1, 1)_{-1}$ & $(0, +1, -1)$ 
& $(+, -, -)$
& $(1, \omega, \omega^{2})$
& $(1, i, -i)$
& $(1, \omega, \underline{\omega})$ \\
\hline
$H 
$ & $(1, 2)_{1/2}$ & $0$ 
& $+$ & $1$ & $1$ & $1$ \\ 
$\Phi 
$ & $(1, 2)_{1/2}$ & $+2$ 
& $+$ & $\omega^{2}$ & $-1$ & $\omega^{2}$ \\ 
$\underline{\Phi} 
$ & $(1, 2)_{1/2}$ & $-2$ 
& $+$ & $\omega$ & $-1$ & $\underline{\omega}^{2}$ \\ 
\hline
\end{tabular}
\caption{Particle contents of models based on $U(1)_{{\mathbf L}_{\mu}-{\mathbf L}_{\tau}}$ 
and $Z_{n}$ flavor symmetries.  The quantum numbers of the SM are also shown in the notation 
of $\big(SU(3)_{c},SU(2)_{L}\big)_{U(1)_{Y}}$. 
For abelian discrete symmetry $Z_{n}$, $\underline{\omega}$ is a conjugate of $\omega$,  
where $\omega$ is $n$-th root of unity. }
\label{Tab:Zn}
\end{table}
Under this flavor symmetry, the following new Yukawa interactions are allowed,
%
\begin{align}
-{\mathcal L}_{U(1)}^\text{yukawa}
=y_{\tau\mu}^{}\, \Phi^{\dag}\, \overline{\tau_{R}^{}}\, L_{\mu}
+\underline{y_{\mu\tau}^{}}\, \underline{\Phi}^{\dag}\, \overline{\mu_{R}^{}}\, L_{\tau} 
+ \text{H.c.}  \label{Eq:U(1)}
\end{align}
%
in addition to the quartic scalar interaction term
$(H^{\dag}\Phi)(H^{\dag}\underline{\Phi})$. 
These interactions easily generate sizable contributions to 
the muon $g-2$ by the scalar mediators as shown in Fig.~\ref{fig:1}. 
In the ordinary gauged $U(1)_{{\mathbf L}_{\mu}-{\mathbf L}_{\tau}}$ model, 
the discrepancy in the muon $g-2$ is explained by 
the new light $Z^\prime$ gauge boson~\cite{Baek:2001kca, Ma:2001md}, 
while in our new proposals a pair of scalar doublets is introduced to give a sizable 
contribution to the muon $g-2$. 
A similar contribution to the muon $g-2$ from the scalar doublets 
are discussed in the model based on the $SU(2)_{\mu\tau}$ symmetry, 
which contains the $U(1)_{{\mathbf L}_{\mu}-{\mathbf L}_{\tau}}$ symmetry as 
a subgroup~\cite{Chiang:2017vcl}. 
In such cases, a pair of scalar doublets plays the primary role in explaining 
the muon $g-2$ anomaly instead of $Z'$ bosons.  
We noted that this new contribution remains even with the unbroken  
$U(1)_{{\mathbf L}_{\mu}-{\mathbf L}_{\tau}}$ flavor symmetry limit. 

From the above consideration in our mind, 
we begin with a global $U(1)_{{\mathbf L}_{\mu}-{\mathbf L}_{\tau}}$ symmetry together with 
a pair of scalar doublets as a simple model for the muon $g-2$ anomaly. 
On the other hand, the $U(1)_{{\mathbf L}_{\mu}-{\mathbf L}_{\tau}}$ symmetry must be broken 
in order to realize observed neutrino masses and mixings~\cite{Asai:2017ryy}. 
If the $U(1)_{{\mathbf L}_{\mu}-{\mathbf L}_{\tau}}$ is not gauged, 
an experimentally unwanted Nambu-Goldstone boson emerges. 
To avoid this problem, we concentrate on abelian discrete symmetries $Z_{n}(n=2, 3, \cdots)$, 
which break the $U(1)_{{\mathbf L}_{\mu}-{\mathbf L}_{\tau}}$ symmetry explicitly.  
\\

The Yukawa interactions based on $Z_{n}$ flavor symmetries are given by 
\begin{align}
-{\mathcal L}_{Z_{2}}^\text{yukawa}
&=
\overline{\ell_{R}^{}} 
\begin{pmatrix} 
y_{e}^{} H^{\dag} +y_{ee}^{}\Phi^{\dag} & & \\
& y_{\mu}^{} H^{\dag} +y_{\mu\mu}^{}\Phi^{\dag} & g_{\mu\tau}^{} H^{\dag} +y_{\mu\tau}^{}\Phi^{\dag} \\
& g_{\tau\mu}^{} H^{\dag} +y_{\tau\mu}^{}\Phi^{\dag} & y_{\tau}^{} H +y_{\tau\tau}^{}\Phi^{\dag}
\end{pmatrix} \!
L 
+ \text{H.c.}  \label{Eq:yZ2} \\
-{\mathcal L}_{Z_{3}}^\text{yukawa}
&=
\overline{\ell_{R}^{}} 
\begin{pmatrix} 
y_{e}^{} H^{\dag} & \underline{y_{e\mu}^{}}\underline{\Phi}^{\dag} & y_{e\tau}^{}\Phi^{\dag} \\
y_{\mu e}^{}\Phi^{\dag} & y_{\mu}^{} H^{\dag} & \underline{y_{\mu\tau}^{}}\underline{\Phi}^{\dag} \\
\underline{y_{\tau e}^{}}\underline{\Phi}^{\dag} & y_{\tau\mu}^{}\Phi^{\dag} & y_{\tau}^{} H^{\dag} 
\end{pmatrix} \!
L 
+ \text{H.c.}  \label{Eq:yZ3}\\
-{\mathcal L}_{Z_{4}}^\text{yukawa}
&=
\overline{\ell_{R}^{}} 
\begin{pmatrix} 
y_{e}^{} H^{\dag} & & \\
& y_{\mu}^{} H^{\dag} & y_{\mu\tau}^{}\Phi^{\dag} \\
& y_{\tau\mu}^{}\Phi^{\dag} & y_{\tau}^{} H^{\dag} 
\end{pmatrix} \!
L
+ \text{H.c.}  \label{Eq:yZ4}\\
-{\mathcal L}_{Z_{n\ge5}}^\text{yukawa}
&=
\overline{\ell_{R}^{}} 
\begin{pmatrix} 
y_{e}^{} H^{\dag} & & \\ 
& y_{\mu}^{} H^{\dag} & \underline{y_{\mu\tau}^{}}\underline{\Phi}^{\dag} \\
& y_{\tau\mu}^{}\Phi^{\dag} & y_{\tau}^{} H^{\dag} 
\end{pmatrix} \!
L
+ \text{H.c.}  \label{Eq:yZ5}
\end{align}
%
The $Z_{n}$ charge assignment in each model is given in Table \ref{Tab:Zn}. 
For $n\ge5$, an accidental global $U(1)_{{\mathbf L}_{\mu}-{\mathbf L}_{\tau}}$ symmetry is recovered 
in the Yukawa interactions taking into account renormalizability. 
Depending on the chosen abelian discrete flavor symmetry, 
a specific structure of the Yukawa interaction is predicted. 
Note that since $\underline{\Phi}$ is identical to $\Phi$ in the $Z_{2}$ and
$Z_{4}$ models, 
the Yukawa interactions of $\underline{\Phi}$ are not shown for these models.  
In the following, we focus on the models with only one extra scalar doublet $\Phi$, 
which minimally explain the muon $g-2$ anomaly. 

In the model based on $Z_{2}$ or $Z_{4}$, possible large new contributions to the muon $g-2$ are retained  
thanks to the existence of the quartic term $(H^{\dag}\Phi)^{2}$. 
From the view of experimental constraints, the $Z_{4}$ model is more
favorable because the $Z_{2}$ model predicts LFV processes 
$\tau\to3\mu, e\mu\mu$ at tree-level, and thus parameter tuning is necessary 
to suppress these processes. 
On the other hand, the LFV processes are automatically forbidden in the (unbroken) $Z_{4}$ model. 
From the view of the numbers of parameters in the model, again the $Z_{4}$ model is preferable 
both in the Yukawa sector and the scalar potential. 
We therefore conclude that the model based on the $Z_{4}$ lepton flavor symmetry is the minimal 
scalar extension of the SM to accommodate the muon $g-2$ anomaly.

\section{The Minimal Model for Muon $g-2$}
\label{sec:2}

 Following the argument in the previous section, we introduce a new
 scalar doublet $\Phi$ to the SM, and impose a $Z_4$ symmetry. 
 The $Z_4$ charge assignment is shown in Table~\ref{Tab:Zn}, and all the
 other fields are trivial under the $Z_{4}$ symmetry.
 The invariant scalar potential is given by
 \begin{align}
  \mathcal{V}
   &=
   \mu_H^2|H|^2+\mu_{\Phi}^{2}|\Phi|^2+\lambda_1|H|^4+\lambda_2|\Phi|^4\nonumber\\
  &~~~+\lambda_3|H|^2|\Phi|^2+\lambda_4|H^{\dag}\Phi|^2
  +\left[\frac{\lambda_5}{2}\left(H^{\dag}\Phi\right)^2+\mathrm{H.c.}\right].
 \end{align}
 This scalar potential is the same as that in the scalar inert doublet model~\cite{Deshpande:1977rw}, 
 where an exact $Z_{2}$ symmetry is preserved in the potential. 
In general, the quartic coupling $\lambda_5$ and the Yukawa couplings $y_{\mu\tau}$,
$y_{\tau\mu}$ are complex. One of the CP phases can be eliminated 
by the field redefinition of $\Phi$. Here, we remove the CP phase
of $\lambda_5$ without loss of generality. 
 Since we demand a stable vacuum, the potential should be bounded from below. 
 The conditions for these requirements are known as ~\cite{Hambye:2009pw}
\begin{align}
 \lambda_1>0,\quad
  \lambda_2>0,\quad
2\sqrt{\lambda_1\lambda_2}+\lambda_3>0,\quad
2\sqrt{\lambda_1\lambda_2}+\lambda_3+\lambda_4\pm|\lambda_5|>0,
\label{eq:p_stab}
\end{align}
at tree level. 
The Higgs doublet $H$ develops a VEV  
as in the SM, and the electroweak symmetry is spontaneously broken. 
The new doublet scalar $\Phi$ is assumed to have a vanishing VEV at leading order.
The scalar fields can then be parameterized  as
 \begin{align}
  H=\left(
     \begin{array}{c}
      0\\
      \left(v+h\right)/\sqrt{2}
     \end{array}
    \right),\qquad
  \Phi=\left(
	\begin{array}{c}
	 \phi^+\\
	 \left(\rho+i\eta\right)/\sqrt{2}
	\end{array}
       \right).
 \end{align}
A component field $h$ corresponds to the Higgs boson with the mass
 $m_h=\sqrt{2\lambda_1}v=125~\mathrm{GeV}$. 
 The electrically neutral component of $\Phi$,  
 $\phi^0=(\rho+i\eta)/\sqrt{2}$, splits into the two mass eigenstates $\rho$ and $\eta$. The masses
 of these neutral states and charged component $\phi^+$ are given by
  \begin{align}
  m_{\rho}^2
   &=
   \mu_\Phi^2+\left(\lambda_3+\lambda_4+\lambda_5\right)\frac{v^2}{2},\\
  m_{\eta}^2
   &=
   \mu_\Phi^2+\left(\lambda_3+\lambda_4-\lambda_5\right)\frac{v^2}{2}, \\
  m_{\phi}^2
   &=
   \mu_\Phi^2+\lambda_3\frac{v^2}{2}. 
  \end{align}
 Thus, one can see that the mass splitting between $\rho$ and $\eta$ is
 controlled by the quartic coupling $\lambda_5$ via the relation
 $m_{\rho}^2-m_{\eta}^2=\lambda_5v^2$. 
\\

\begin{figure}[t]
\centering
   \includegraphics[scale=1.1]{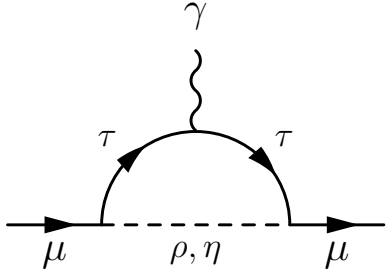}
   \caption{Feynman diagram inducing muon anomalous magnetic moment in
   $Z_4$ model.}
   \label{fig:2}
 \end{figure}

 In this model, the new contribution to muon anomalous magnetic moment
 comes from Fig.~\ref{fig:2}, which is computed as 
\begin{align}
 \Delta a_{\mu}^\mathrm{new}
  &=
  \frac{\mathrm{Re}\left(y_{\mu\tau}y_{\tau\mu}\right)}{(4\pi)^2}
  \left[\frac{m_\mu
   m_\tau}{m_\rho^2} I_1(m_\mu^2/m_\rho^2, m_\tau^2/m_\rho^2)
   -\frac{m_\mu
   m_\tau}{m_\eta^2} I_1(m_\mu^2/m_\eta^2, m_\tau^2/m_\eta^2) \right]\nonumber\\
 &~~~+\frac{|y_{\mu\tau}|^2+|y_{\tau\mu}|^2}{2(4\pi)^2} \left[ 
 \frac{m_\mu^2}{m_\rho^2} I_2(m_\mu^2/m_\rho^2, m_\tau^2/m_\rho^2) 
 +\frac{m_\mu^2}{m_\eta^2} I_2(m_\mu^2/m_\eta^2, m_\tau^2/m_\eta^2) \right],
\label{eq:g-2}
\end{align} 
where the loop functions $I_1(a,b)$ and $I_2(a,b)$ are defined by
\begin{align}
 I_1(a,b)
  &\equiv
  \int_0^1\frac{(1-x)^2}{x-x(1-x)a+(1-x)b}dx,\\
 I_2(a,b)
  &\equiv
  \frac{1}{2}\int_0^1\frac{x(1-x)^2}{x-x(1-x)a+(1-x)b}dx.
\end{align}
Note that the contribution in the first line of Eq.~\eqref{eq:g-2} is
dominant compared to that in the second line with an enhancement factor  
$m_{\tau}/m_{\mu}\approx17$, because of the chirality flipping effect. 
The numerical value of these loop functions are always positive, and thus
the sign of the new contribution is determined by the relative sign of 
$\mathrm{Re}\left(y_{\mu\tau}y_{\tau\mu}\right)$ and $m_\rho^2-m_{\eta}^2=\lambda_5v^2$.

In the numerical analysis, we require that the discrepancy of muon $g-2$
is improved to be within $2\sigma$ range after including the new physics contribution. 
Thus, $\Delta a_{\mu}^\mathrm{new}$ in Eq.~\eqref{eq:g-2} should
be in the interval~\cite{Chiang:2017vcl} 
\begin{align}
115\times10^{-11}<\Delta a_{\mu}^\mathrm{new}<421\times10^{-11}.
\end{align}

\section{The Constraints}
 \subsection{Electroweak Precision Tests}
 The new scalar particles $\rho$, $\eta$ and $\phi^+$ affect the electroweak
 precision observables through vacuum polarization diagrams. 
 These are conveniently parameterized by the so-called $S, T, U$-parameters~\cite{Peskin:1991sw}. 
 The expression of the $S, T, U$-parameters in this model is the same as 
 that in the inert doublet model~\cite{Barbieri:2006dq} or 
 in the THDM~\cite{Peskin:2001rw, Kanemura:2011sj} with the alignment limit, which are
 given by 
\begin{align}
 S&=
  \frac{1}{2\pi}\biggl[\frac{1}{12}\log\frac{m_\rho^2m_\eta^2}{m_\phi^4}+G\left(m_\rho^2,m_\eta^2\right)\biggr],\\
 T&=
  \frac{\sqrt{2}G_{\!F}}{(4\pi)^2\alpha_\mathrm{em}}
  \Bigl[
   F\left(m_{\phi}^2,m_\rho^2\right)+F\left(m_{\phi}^2,m_\eta^2\right)-F\left(m_\rho^2,m_\eta^2\right)
   \Bigr],\\
 U&=
  \frac{1}{2\pi}
  \Bigl[
  G\left(m_{\phi}^2,m_\rho^2\right)+G\left(m_{\phi}^2,m_\eta^2\right)-G\left(m_\rho^2,m_\eta^2\right)
\Bigr],
\end{align}
where $G_{\!F}$ is the Fermi constant, $\alpha_\mathrm{em}$ is the
electromagnetic fine structure constant, and the functions $F(x,y)$ and $G(x,y)$ are given by
\begin{align}
 F(x,y)&=
  \frac{x+y}{2}-\frac{xy}{x-y}\log\left(\frac{x}{y}\right),\\
 G(x,y)&=
  -\frac{5x^2-22xy+5y^2}{36(x-y)^2}+\frac{x^3-3x^2y-3xy^2+y^3}{12(x-y)^3}\log\left(\frac{x}{y}\right).
\end{align}
The current experimental bounds on these parameters are summarized as~\cite{Baak:2014ora}
\begin{align}
 S=0.05\pm0.11,\quad
  T=0.09\pm0.13,\quad
  U=0.01\pm0.11,
\end{align}
with correlation coefficients $0.90$ between $S$ and $T$, $-0.59$
between $S$ and $U$, and $-0.83$ between $T$ and $U$, respectively. 

We impose the requirement that the theoretical prediction on these
parameters should be kept in  the $2\sigma$ range of the experimental values. 
If relatively light new particles $(\lesssim m_{W})$ are mediated in a loop, 
more sophisticated analysis of the electroweak precision tests may be applied 
as in a lepton-specific THDM~\cite{Abe:2017jqo}.

\subsection{Lepton Universality in Charged Lepton Decays}
\begin{figure}[t]
\centering
 \includegraphics[scale=0.85]{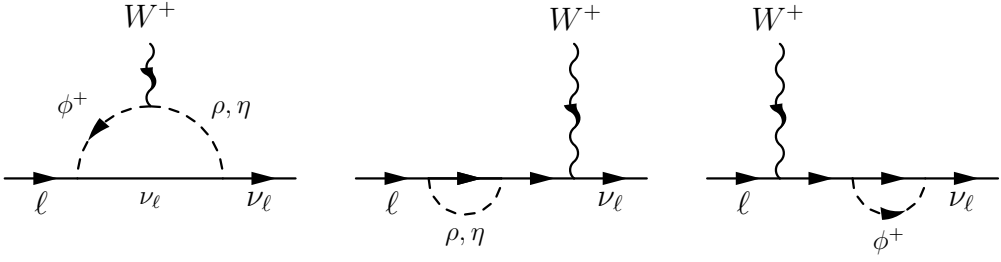}
 \caption{Feynman diagrams of the loop corrections to charged lepton currents $-gW_{\mu}^+\overline{\nu_{\ell_L}}\gamma^{\mu}\ell_L$.}
 \label{fig:uni}
\end{figure}
The new Yukawa couplings $y_{\mu\tau}$ and $y_{\tau\mu}$ give additional contributions 
to the decay of charged leptons. 
First, the new tau decay mode $\tau\to\mu\,\overline{\nu_\tau^{}}\, \nu_\mu^{}$
is induced at tree level. 
The partial decay width is calculated as
\begin{align}
 \Gamma_{\tau\to\mu\,\overline{\nu_\tau^{}}\,\nu_\mu^{}}=
  \frac{|y_{\mu\tau}|^2|y_{\tau\mu}|^2m_{\tau}^5}{6144\pi^3m_{\phi}^4}\, f(m_\mu^2/m_\tau^2)\,  
  r_W^{\tau}r_\gamma^{\tau},
\label{eq:lep_uni}
\end{align}
where $f(x)$ is the kinematic function $f(x)=1-8x+8x^3-x^4-12x^2\log{x}$, 
the factor $r_W^\tau$ is the $W$-boson propagator correction, 
and $r_\gamma^\tau$ is the QED radiative correction, which are given
by~\cite{Marciano:1988vm}
\begin{align}
 r_W^\ell\equiv1+\frac{3m_\ell^2}{5m_W^2},\quad
  r_\gamma^{\ell}\equiv1+\frac{\alpha_\mathrm{em}(m_\ell)}{2\pi}\left(\frac{25}{4}-\pi^2\right). 
\end{align}
In the numerical evaluation, we use PDG data for the $W$ boson mass, charged lepton masses, 
the electromagnetic fine structure constant~\cite{Tanabashi:2018oca}. 
If we worked out in the mass eigenbasis of neutrinos, 
one may expect interference effect in $\tau\to\mu\,\overline{\nu_i^{}}\,\nu_j^{}\, (i=1,2,3)$. 
Such effect is, however, negligible since the chirality flip occurs and
it is suppressed by small neutrino masses. 

Second, one-loop corrections in the charged lepton currents are induced 
by new Yukawa interactions as shown in Fig.~\ref{fig:uni}. 
Although each diagram includes a divergence, it cancels out after the sum over all the graphs. 
We then obtain a finite correction without renormalization. 
Following the results in Ref.~\cite{Abe:2015oca} for the Type-X THDM, 
we define the loop corrections $\delta
g_{W\overline{\nu_\ell}^{}\ell}$ as 
$g\to g\left(1+\delta g_{W\overline{\nu_\ell}\ell}^{}\right)$. 
The results in the $\mu$-$\tau$-specific scalar doublet model are  
\begin{align}
\delta
 g_{W\overline{\nu_\mu}\mu}^{} 
 =\frac{|y_{\tau\mu}|^2}{2(4\pi)^2}\, I_L(m_\rho^2/m_{\phi}^2, m_\eta^2/m_{\phi}^2),\quad
 \delta g_{W\overline{\nu_\tau}\tau}^{}
 =\frac{|y_{\mu\tau}|^2}{2(4\pi)^2}\, I_L(m_\rho^2/m_{\phi}^2, m_\eta^2/m_{\phi}^2),
\label{eq:lep_uni2}
\end{align}
where the small lepton masses are neglected, and 
the loop function $I_L(x,y)$ is defined by 
\begin{align}
I_L(x,y)\equiv1+\frac{1}{4}\frac{1+x}{1-x}\log{x}+\frac{1}{4}\frac{1+y}{1-y}\log{y}.
\end{align}

Taking into account above corrections at tree level and the one-loop level, 
the total leptonic decay widths of muon and tau lepton are summarized as 
\begin{align}
 \Gamma_{\tau\to\mu\overline{\nu}\nu}
  &\equiv
  \Gamma_{\tau\to\mu\overline{\nu_\mu}\nu_{\tau}}
  +\Gamma_{\tau\to\mu\overline{\nu_\tau}\nu_{\mu}}\nonumber\\
  &=
  \frac{m_\tau^5}{6144\pi^3}\left[\frac{g^4\left(1+\delta
 g_{W\overline{\nu_\tau}\tau}^{}\right)^2\left(1+\delta
 g_{W\overline{\nu_\mu}\mu}^{}\right)^2}{m_W^4}+\frac{|y_{\mu\tau}|^2|y_{\tau\mu}|^2}{m_\phi^4}\right] 
  f(m_\mu^2/m_\tau^2)\, r_W^{\tau}r_\gamma^{\tau},\label{eq:tau_decay}\\
 \Gamma_{\tau\to e\overline{\nu}\nu}
  &\equiv
  \Gamma_{\tau\to e\overline{\nu_e}\nu_{\tau}}
  =
  \frac{g^4\left(1+\delta g_{W\overline{\nu_\tau}\tau}^{}\right)^2m_\tau^5}{6144\pi^3m_W^4}\, 
  f(m_e^2/m_\tau^2)\, r_W^{\tau}r_\gamma^{\tau},\\
 \Gamma_{\mu\to e\overline{\nu}\nu}
  &\equiv
  \Gamma_{\mu\to e\overline{\nu_e}\nu_{\mu}}
  =
  \frac{g^4\left(1+\delta g_{W\overline{\nu_\mu}\mu}^{}\right)^2m_\mu^5}{6144\pi^3m_W^4}\, 
  f(m_e^2/m_\mu^2)\, r_W^{\mu}r_\gamma^{\mu}.
\end{align}
In Eq.~(\ref{eq:tau_decay}), the decay widths for the channels
$\tau\to\mu\,\overline{\nu_{\mu}^{}}\,\nu_\tau^{}$ and
$\tau\to\mu\,\overline{\nu_{\tau}^{}}\,\nu_\mu^{}$ are combined since these processes cannot be 
distinguished in actual measurements.
In general, the above leptonic decay widths are conveniently parameterized as 
\begin{align}
\Gamma_{\ell\to\ell^\prime\overline{\nu}\nu}
=
 \frac{G_{\ell}G_{\ell^\prime}m_\ell^5}{192\pi^3}\, 
 f(m_{\ell^\prime}^2/m_\ell^2)\, r_W^\ell r_\gamma^\ell,
\end{align}
with $G_\ell\equiv g_\ell^2/\left(4\sqrt{2}m_W^2\right)$.
The effective weak couplings for leptons $g_\ell~(\ell=e,\mu,\tau)$ are
severely constrained as~\cite{Amhis:2014hma}
\begin{align}
 \frac{g_{\tau}}{g_{\mu}}=1.0011\pm0.0015,\quad
  \frac{g_{\tau}}{g_{e}}=1.0029\pm0.0015,\quad
  \frac{g_{\mu}}{g_{e}}=1.0018\pm0.0014,
\end{align}
with correlation coefficients $0.53$ between
$g_\tau/g_\mu$ and $g_\tau/g_e$, $-0.49$ between
$g_\tau/g_\mu$ and $g_\mu/g_e$, and $0.48$ between
$g_\tau/g_e$ and $g_{\mu}/g_e$, respectively. 
Using this notation, we find analytic expressions for the corresponding quantities, 
as   
\begin{align}
   \frac{g_\tau}{g_\mu}
   &=\frac{1+\delta g_{W\overline{\nu_\tau}\tau}^{}}{1+\delta
   g_{W\overline{\nu_\mu}\mu}^{}},\\
   \frac{g_\tau}{g_e}
   &=\left(1+\delta
   g_{W\overline{\nu_\tau}\tau}^{}\right)\sqrt{1+\frac{m_W^4|y_{\mu\tau}|^2|y_{\tau\mu}|^2}{g^4\left(1+\delta
   g_{W\overline{\nu_\mu}\mu}^{}\right)^2\left(1+\delta g_{W\overline{\nu_\tau}\tau}^{}\right)^2m_\phi^4}},\\
   \frac{g_\mu}{g_e}
   &=\left(1+\delta
   g_{W\overline{\nu_\mu}\mu}^{}\right)\sqrt{1+\frac{m_W^4|y_{\mu\tau}|^2|y_{\tau\mu}|^2}{g^4\left(1+\delta
   g_{W\overline{\nu_\mu}\mu}^{}\right)^2\left(1+\delta g_{W\overline{\nu_\tau}\tau}^{}\right)^2m_\phi^4}}.
  \end{align}
In the numerical analysis, we demand that these quantities should be 
in the $2\sigma$ range of the experimental values.

\subsection{Lepton Universality in $Z$ Boson Decays}

\begin{figure}[t]
\centering
 \includegraphics[scale=0.80]{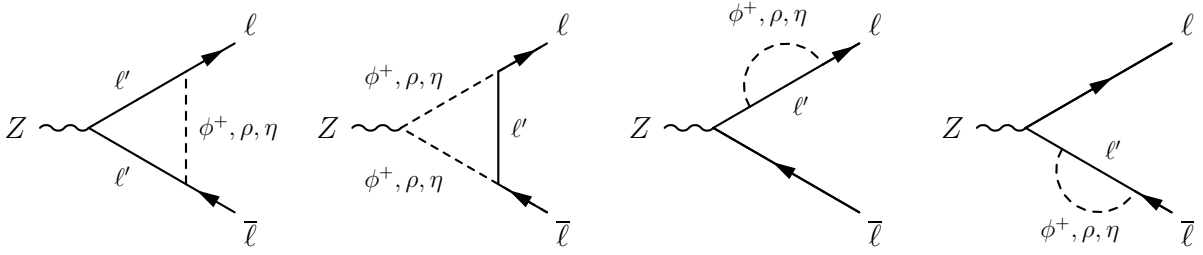}
 \caption{Feynman diagrams of the loop corrections to $Z$ boson leptonic decays.}
 \label{fig:z_decay}
\end{figure}

The $Z$ boson leptonic decays are also modified by the Yukawa couplings.
In general, the interactions between $Z$ boson and a pair of charged leptons can be written as 
\begin{align}
 \mathcal{L}
 =-\frac{g}{\cos\theta_W^{}}Z_{\mu}\, 
 \overline{\ell}\gamma^{\mu} (g_L^{\ell}P_L+g_R^{\ell}P_R)\ell,
\end{align}
where $g_L^\ell$ and $g_R^\ell$ are given by
$g_L^{\ell}=-1/2+\sin^2\theta_W^{}$ and $g_R^{\ell}=\sin^2\theta_W^{}$ at tree
level, which are universal over lepton flavors. 
With this convention, the leptonic decay widths are calculated as
\begin{align}
 \Gamma\left(Z\to\ell\overline{\ell}\right) 
 =\frac{g^2m_Z}{24\pi\cos^2\theta_W^{}}\left(|g_L^{\ell}|^2+|g_R^{\ell}|^2\right),
\end{align}
where $\theta_W$ is the Weinberg angle. 
The couplings $g_L^\ell$ and $g_R^\ell$ receive the one-loop corrections 
from the diagrams shown in Fig.~\ref{fig:z_decay}. 
 The total one-loop correction is finite while each diagram includes a
 divergence as same as the case of the charged lepton decay vertices. 
The loop corrections for the neutral current interaction with tau lepton defined by $g_{L/R}^{\tau}\to
 g_{L/R}^{\tau}+\delta g_{L/R}^{\tau}$ are parameterized as~\cite{Chun:2016hzs}
\begin{align}
 \delta g_L^\tau=a_L^{\tau}+\sin^2\theta_W^{} b_L^{\tau},\qquad
  \delta g_R^\tau=a_R^{\tau}+\sin^2\theta_W^{} b_R^{\tau}.
\end{align}
Neglecting the small lepton masses, the coefficients $a_L^{\tau}$,
$b_L^{\tau}$, $a_R^{\tau}$ and $b_R^{\tau}$ are computed as
\begin{align}
 a_L^{\tau} &=
  \frac{|y_{\mu\tau}|^2}{2(4\pi)^2}
  \left[
   -\frac{1}{2}B_Z(\xi_\rho)-\frac{1}{2}B_Z(\xi_\eta)-2C_Z(\xi_{\rho},\xi_{\eta})
	    \right],\label{eq:aL}\\
 b_L^{\tau} &=
  \frac{|y_{\mu\tau}|^2}{2(4\pi)^2}
  \left[
   B_Z(\xi_\rho)+B_Z(\xi_\eta)+\tilde{C}_Z(\xi_\rho)+\tilde{C}_Z(\xi_\eta)
		      \right],\label{eq:bL}\\
 a_R^{\tau} &=
  \frac{|y_{\tau\mu}|^2}{2(4\pi)^2}
  \left[
   2C_Z(\xi_\rho,\xi_\eta)-2C_Z(\xi_\phi,\xi_\phi)+\tilde{C}_Z(\xi_\phi)-\frac{1}{2}\tilde{C}_Z(\xi_\rho)-\frac{1}{2}\tilde{C}_Z(\xi_\eta)
	    \right],\label{eq:aR}\\
 b_R^{\tau} &=
  \frac{|y_{\tau\mu}|^2}{2(4\pi)^2}
  \left[
   B_Z(\xi_\rho)+B_Z(\xi_\eta)+2B_Z(\xi_\phi)+\tilde{C}_Z(\xi_\rho)+\tilde{C}_Z(\xi_\eta)+4C_Z(\xi_\phi,\xi_\phi)
		      \right],\label{eq:bR}
\end{align}
where $\xi_a\equiv m_a^2/m_Z^2~(a=\phi,\rho,\eta)$, and the loop functions $B_Z(\xi)$,
$\tilde{C}_Z(\xi)$ and $C_Z(\xi_1,\xi_2)$ 
are defined by~\cite{Chun:2016hzs}
\begin{align}
 B_Z(\xi)
  &\equiv
  -\frac{1}{4}+\frac{1}{2}\log\xi,\\
 \tilde{C}_Z(\xi)
  &\equiv
  \frac{1}{2}-\xi\left(1+\log\xi\right)+\xi^2\left[\log\xi\log\left(\frac{1+\xi}{\xi}\right)-\mathrm{Li}_{2}\left(-\frac{1}{\xi}\right)\right]\nonumber\\
 &\hspace{0.4cm}
  -\frac{i\pi}{2}\left[1-2\xi+2\xi^2\log\left(\frac{1+\xi}{\xi}\right)\right],\\
 C_Z(\xi_1,\xi_2)
  &\equiv
  -\frac{1}{2}\lim_{\epsilon\to0}\int_{0}^{1}dx\int_{0}^{1-x}dy\log\left(x\xi_1+y\xi_2-xy-i\epsilon\right).
\end{align}
Similarly, the loop correction with muon is obtained by replacing
$y_{\mu\tau}\leftrightarrow y_{\tau\mu}$ in Eq.~(\ref{eq:aL})-(\ref{eq:bR}), 
and there is no loop correction for the neutral current interaction with electron at this order.

The ratios of the $Z$ boson leptonic decays are constrained by LEP~\cite{ALEPH:2005ab} 
\begin{equation}
 \frac{\Gamma\left(Z\to
	      \mu\overline{\mu}\right)}{\Gamma\left(Z\to e\overline{e}\right)}=1.0009\pm0.0028,\qquad
\frac{\Gamma\left(Z\to \tau\overline{\tau}\right)}{\Gamma\left(Z\to e\overline{e}\right)}=1.0019\pm0.0032,
\end{equation}
with correlation coefficient $0.63$. 
We require that these ratios take values in the $2\sigma$ range of the LEP data in the numerical study.

\subsection{Collider Limits}
The lower bound of the charged scalar mass is given as
$m_\phi\gtrsim93.5~\mathrm{GeV}$ by LEP~\cite{Abbiendi:2013hk}. 
There are also LHC bounds which depend on branching ratio of the charged scalar $\phi^+$. 
Since the charged scalar in our model has the same quantum charges with
the charged sleptons in supersymmetric models except for the matter parity, 
the bound for sleptons from the electroweak production can be applied for 
$\phi^{+}$ if the dominant (prompt) decay channels are  
$\phi^+\to\overline{\tau}\,\nu_\mu^{},\overline{\mu}\,\nu_\tau^{}$. 
The slepton mass bound in the massless neutralino limit 
can be recast to $m_\phi\gtrsim700~\mathrm{GeV}$~\cite{Aaboud:2018jiw, ATLAS:2019cfv}.~\footnote{
In the Ref.~\cite{ATLAS:2019cfv}, the bound is obtained assuming three generations of 
mass-degenerate left- and right-handed sleptons. 
On the other hand, the charged scalar in our model corresponds to a single generation 
of a left-handed slepton, which equally decays into $\mu\nu$ and $\tau\nu$. 
Therefore, the bound $m_\phi\gtrsim700~\mathrm{GeV}$ seems to be a conservative estimate. 
}
On the other hand,  if $\phi^+$ is heavier than $\rho$ or $\eta$ 
the decay channels $\phi^+\to W^+\rho, W^+\eta$ open and can be dominant.
In such a case, the mass bound for sleptons cannot be simply applied, 
and $m_{\phi}$ can be lighter than $700~\mathrm{GeV}$. 
Thus, we choose $m_{\phi} = 200$ and $700$ GeV as representative values in the numerical analysis.

\subsection{Triviality Bound}
Even if the couplings in the model are perturbative at electroweak
scale, it may become non-perturbative at a high energy scale
after including renormalization group running of the couplings. 
In particular, if the couplings are $\mathcal{O}(1)$ at electroweak
scale, it can quickly increase, and tends to become non-perturbative 
around $\mathcal{O}(10-100)~\mathrm{TeV}$. 
The $\beta$ functions for the renormalization group running at one loop level 
are collected in Appendix A, where the SM Yukawa couplings are neglected 
except for the top Yukawa coupling $y_t$. 
We solve the coupled renormalization group equations from the $Z$ boson
 mass scale to the cut-off scale $\Lambda$. In the numerical analysis,
 we take $\Lambda=100~\mathrm{TeV}$. 
 Then, we demand that all the couplings in the model are perturbative until the cut-off scale. 
 Namely, the required conditions are: $|\lambda_i|\leq
 4\pi~(i=1-5)$ and $|y_t| ,|y_{\mu\tau}|,|y_{\tau\mu}|\leq\sqrt{4\pi}$ at the cut-off scale. 

\subsection{Numerical Analysis}

\begin{figure}[t]
\centering
 \includegraphics[scale=0.7]{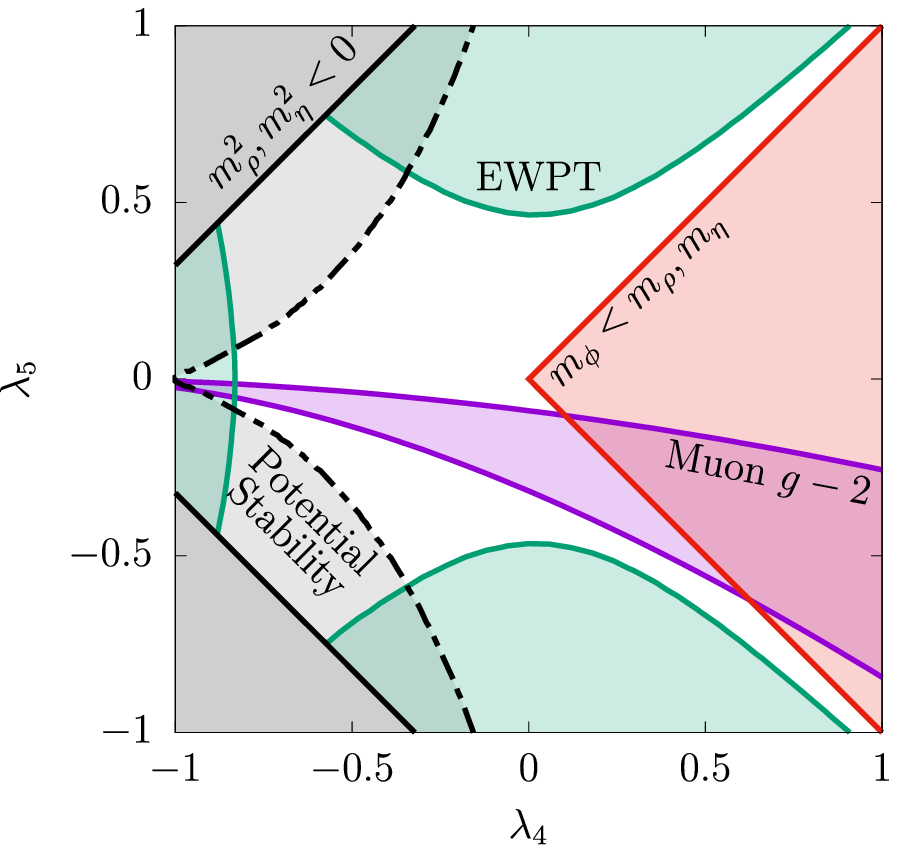}
 \includegraphics[scale=0.7]{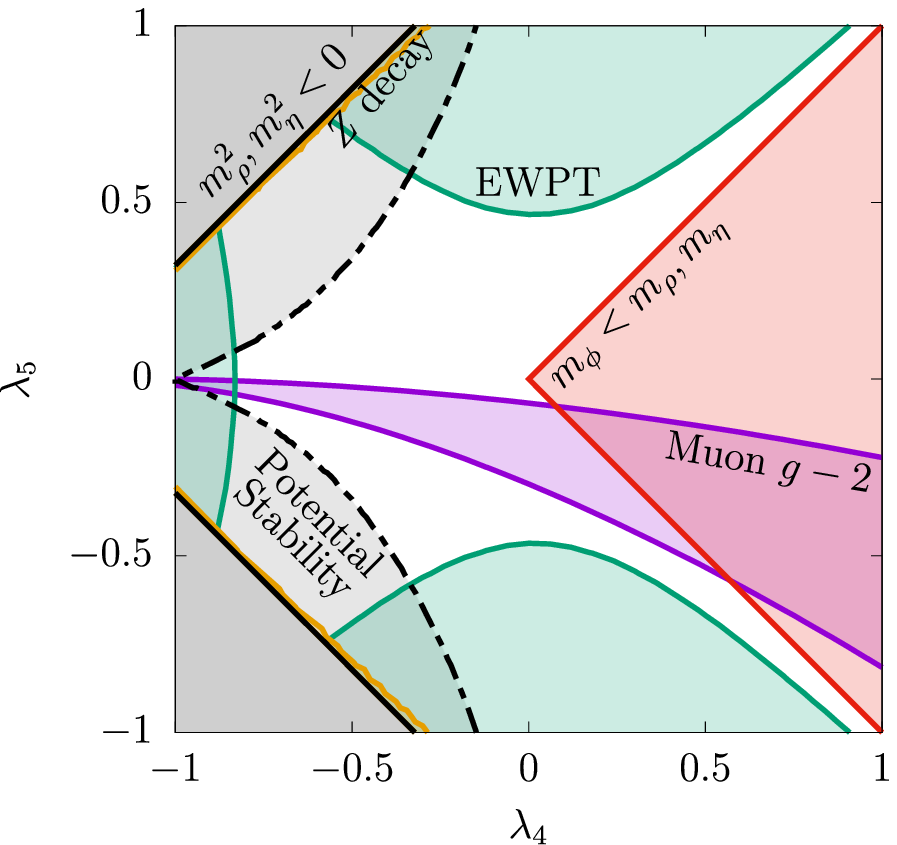}\\
 \hspace{0.2cm}
 \includegraphics[scale=0.7]{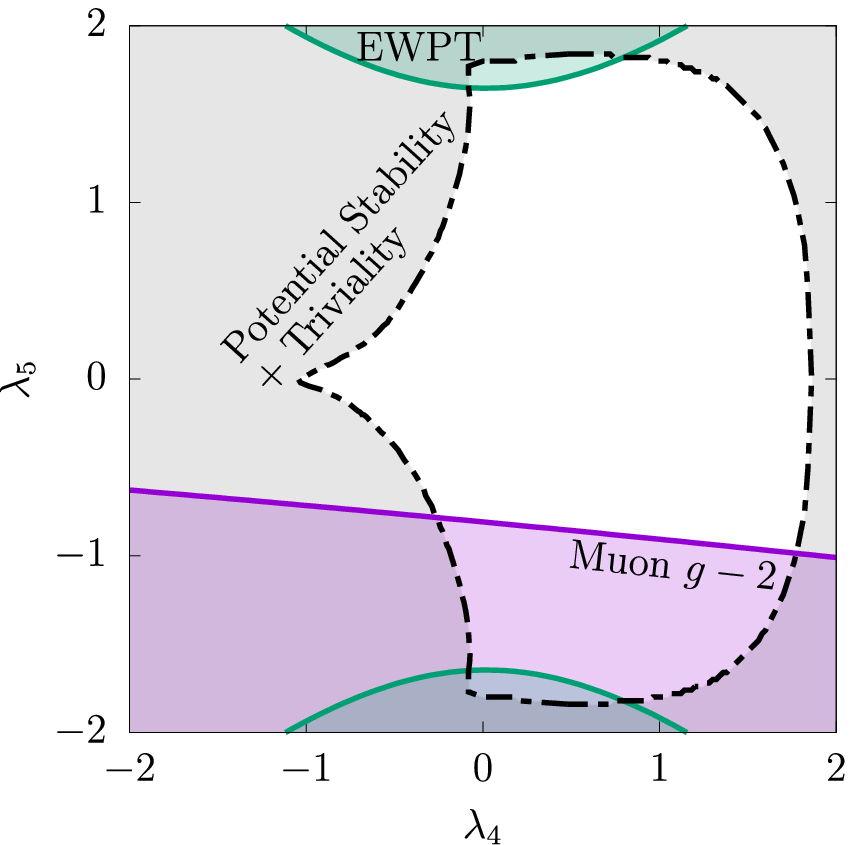}
 \hspace{0.2cm}
 \includegraphics[scale=0.7]{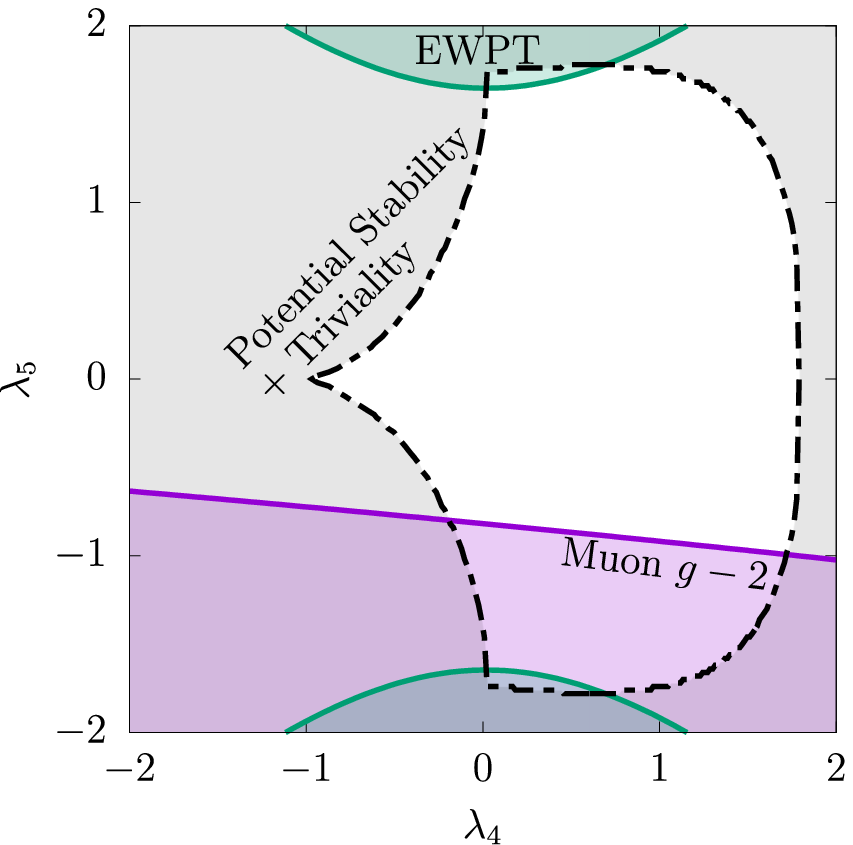}
 \caption{Numerical analysis in ($\lambda_4$, $\lambda_5$) plane, 
 where the charged scalar mass and Yukawa couplings 
 ($m_\phi$, $y_{\mu\tau}$, $y_{\tau\mu}$) are fixed as 
 $(200~\mathrm{GeV}$, $0.20$, $0.20$) on top left panel, 
 ($200~\mathrm{GeV}$, $0.04$, $1.00$) on top right panel, 
 ($700~\mathrm{GeV}$, $0.70$, $0.70$) on bottom left panel, and
 ($700~\mathrm{GeV}$, $0.41$, $1.20$) on bottom right panel,
 respectively.
 }
\label{fig:num1}
\end{figure}

\begin{figure}[t]
\centering
  \includegraphics[scale=0.7]{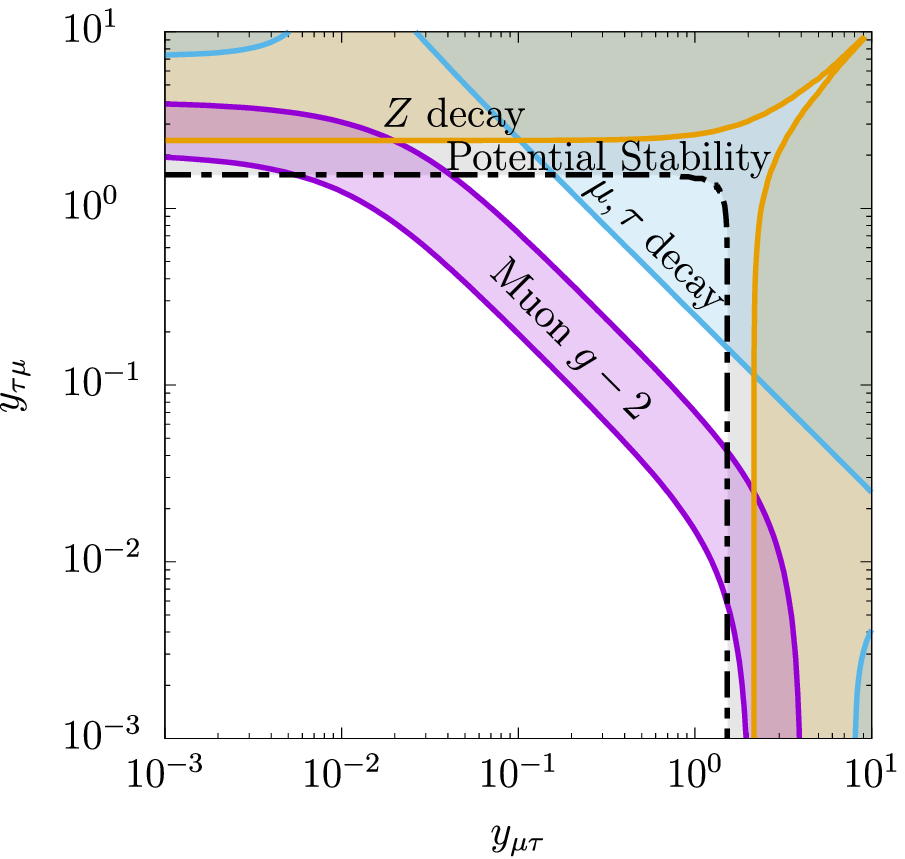}
  \includegraphics[scale=0.7]{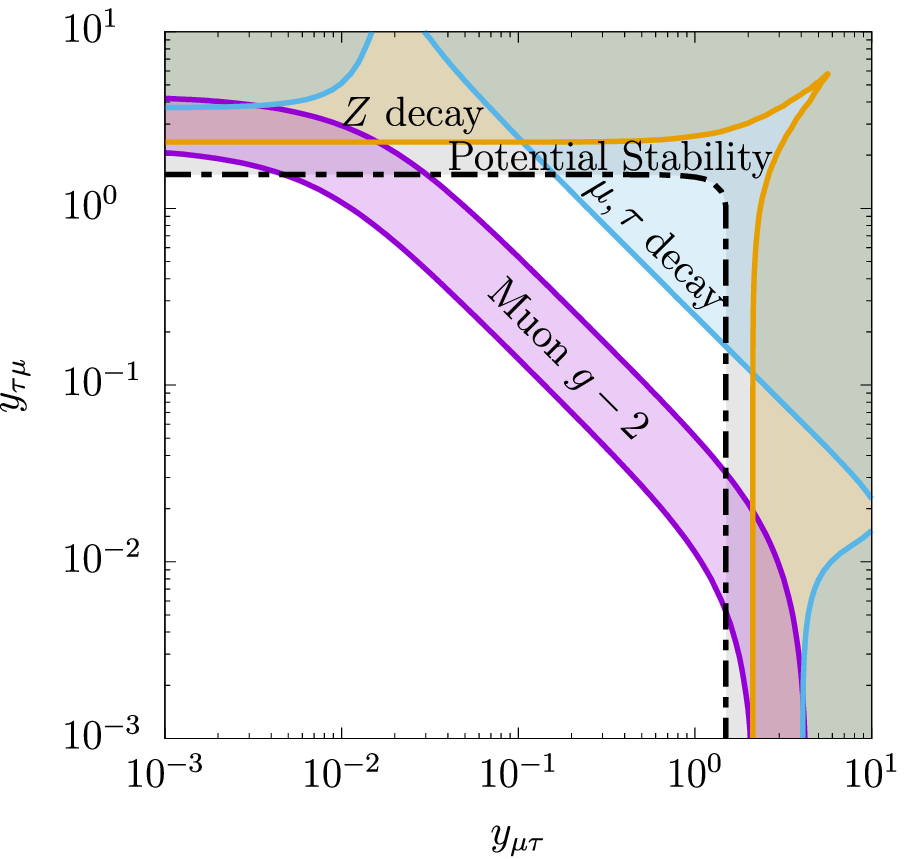}\\
  \includegraphics[scale=0.7]{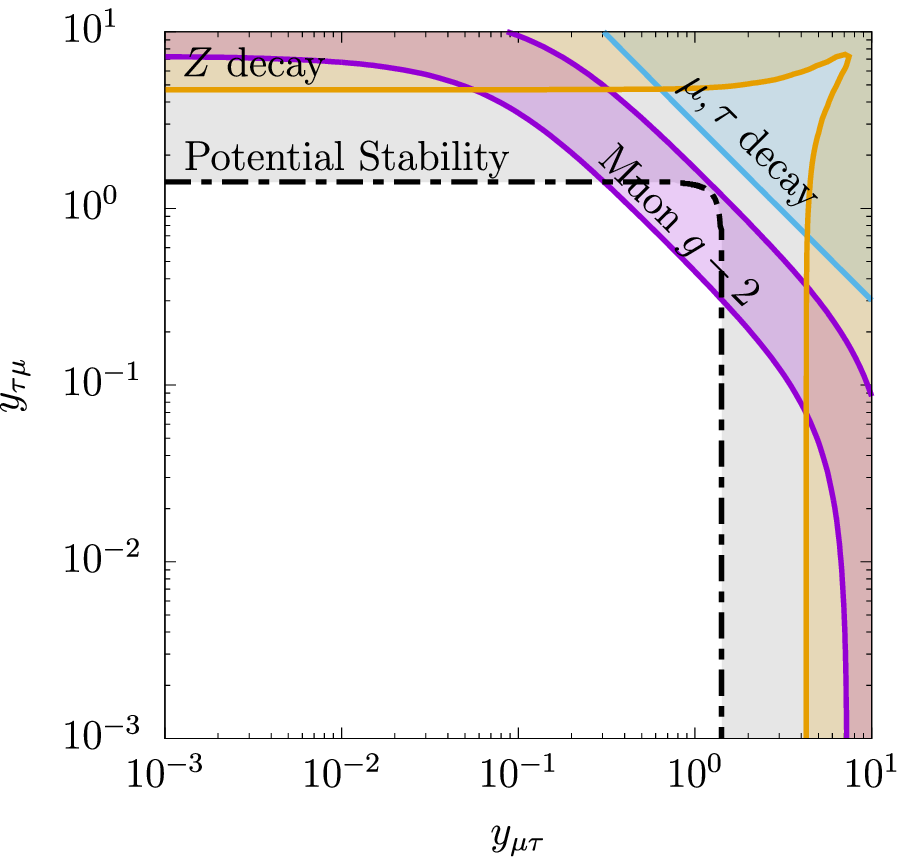}
  \includegraphics[scale=0.7]{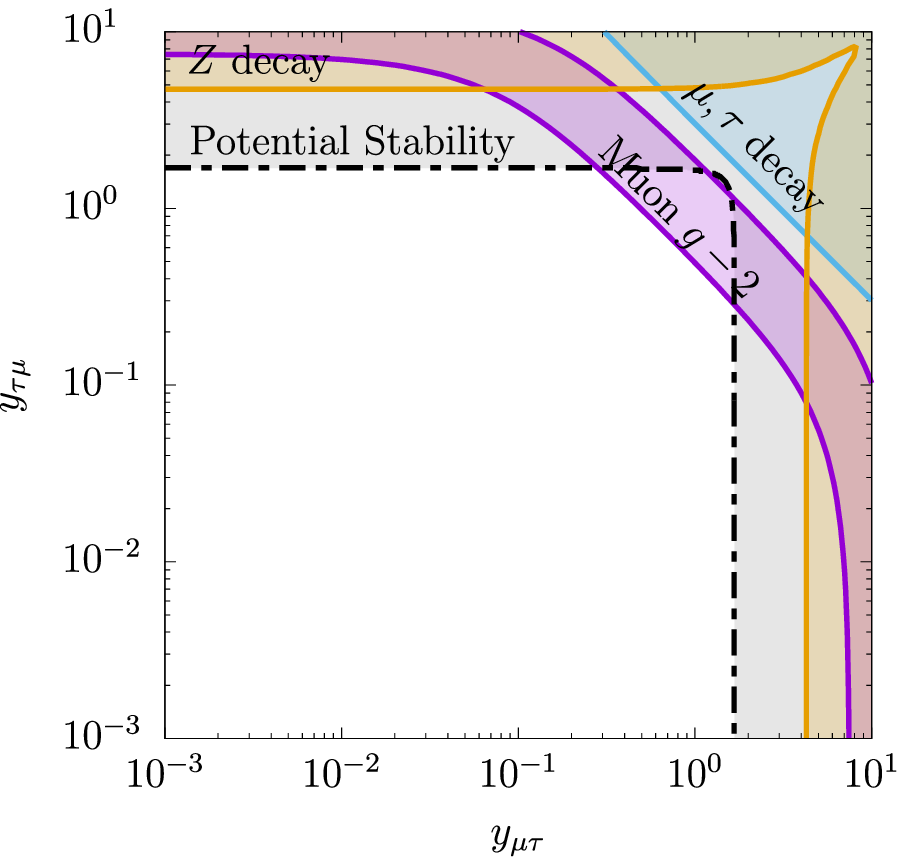}
  \caption{Numerical analysis in ($y_{\mu\tau}$, $y_{\tau\mu}$) plane, 
  where the charged scalar mass and the scalar quartic couplings
  ($m_\phi$, $\lambda_4$, $\lambda_5$) are fixed as
  ($200~\mathrm{GeV}$, $-0.01$, $-0.20$) on top left panel,
  ($200~\mathrm{GeV}$, $0.30$, $-0.40$) on top right panel,
  ($700~\mathrm{GeV}$, $0.01$, $-1.00$) on bottom left panel and
  ($700~\mathrm{GeV}$, $1.00$, $-1.00$) on bottom right panel,
 respectively.
 }
  \label{fig:num2}
\end{figure}

We explore parameter space which can explain the discrepancy in the muon $g-2$  
while satisfying the relevant constraints. 
In Fig.~\ref{fig:num1}, we present the numerical analysis 
in the ($\lambda_4$, $\lambda_5$) plane for fixed values of charged scalar mass  
($m_{\phi}$) and Yukawa couplings ($y_{\mu\tau}$, $y_{\tau\mu}$). 
In this subsection, we restrict new Yukawa couplings to be real. 
The upper (lower) two panels show the low (high) mass scenarios with $m_{\phi}=200\, (700)$ GeV. 
In the left panels, we maximize the new physics contributions to the muon $g-2$, 
where $y_{\mu\tau}=y_{\tau\mu}$ is assumed, while hierarchical Yukawa
couplings are taken in the right panels such that the magnitude of the product
$y_{\mu\tau}y_{\tau\mu}$ is retained as same with the left panels so
that the parameter space favored by muon $g-2$ does not change. 
The purple region represents the parameter space which can accommodate
the muon magnetic moment anomaly at $2\sigma$ confidence level (CL). 
In the top panels in Fig.~\ref{fig:num1}, the left-top and left-bottom
region colored by gray is forbidden because the mass of the 
neutral scalars $\rho$ or $\eta$ becomes negative. 
The green region is ruled out by the electroweak precision tests at
$2\sigma$ CL. 
This constraint becomes stronger for lighter scalar masses. 
On the other hand, even if the charged scalar mass is relatively light, 
the constraint can be evaded if $\lambda_4\sim\pm\lambda_5$,
which implies that one of $\rho$ and $\eta$ is nearly degenerate with
the charged scalar $\phi^+$. 
The orange region is excluded by the constraint of the lepton universality
($Z$ boson decays). 
Since the loop corrections to the $Z$ boson decays given by
Eq.~(\ref{eq:aL})-(\ref{eq:bR}) are proportional to $|y_{\mu\tau}|^2$ or
$|y_{\tau\mu}|^2$, one can find that the constraint becomes stronger for
larger hierarchy between $y_{\mu\tau}$ and $y_{\tau\mu}$ for a fixed
$y_{\mu\tau}y_{\tau\mu}$. 
Note that the loop corrections for the charged lepton currents given by
Eq.~(\ref{eq:lep_uni2}) also have the same dependence on the Yukawa couplings. 
However, the constraint from muon and tau lepton decays are
slightly weaker than the $Z$ boson decays in the above parameter sets. 
The red region shows the parameter space that the charged scalar
$\phi^+$ becomes the lightest than the neutral scalars $\rho$ and $\eta$. 
In this region, since the charged scalar decays dominantly into
a pair of a charged lepton and a neutrino, the LHC mass limit ($m_\phi\gtrsim700~\mathrm{GeV}$) is applied. 
The outside of the dot-dashed curve colored by gray is disfavored  
by the potential stability conditions given by Eq~(\ref{eq:p_stab}) and the triviality. 
The negative $\lambda_4$ region tends to be excluded by the
potential stability conditions while the
remaining region is bounded by the triviality of the quartic
couplings $\lambda_i~(i=1-5)$. 
Here, we take $\lambda_2=\lambda_3=0.5$ at the $Z$ boson mass scale as
an initial condition of the renormalization group equation. 
Note that if smaller couplings $\lambda_2$ and $\lambda_3$ are assumed, the bound of the potential stability 
becomes stronger as we expect from Eq.~(\ref{eq:p_stab}).

In Fig.~\ref{fig:num2}, we show the parameter space in the ($y_{\mu\tau}$, $y_{\tau\mu}$) plane 
by fixing the scalar masses $m_{\rho}$, $m_{\eta}$ and $m_\phi$. 
The positive Yukawa coupling $y_{\mu\tau}$ is chosen without loss
of generality.
We here concentrate on the case with negative values of $\lambda_5$,
which is favored by the muon $g-2$ anomaly together with a positive
value of $y_{\tau\mu}$.
At the same time, we assume a negative $\lambda_{4}+\lambda_{5}$ for $m_\phi=200~\mathrm{GeV}$. 
This parameter choice allows the cascade decay of the charged scalar to other scalars,
and therefore we can avoid the strong constraint on the charged scalar mass
from the LHC slepton search.
The purple region can accommodate the muon $g-2$ anomaly at $2\sigma$ CL, 
while the orange and light blue region are excluded by the lepton
universality of the $Z$ boson decays and the charged lepton decays, respectively. 
The constraint of the electroweak precision tests is satisfied in all the plots, 
which does not depend on the Yukawa couplings. 
One can see from Fig.~\ref{fig:num2} that the constraint of the charged
lepton decays (light blue) is always stronger than that of the $Z$ boson
decay (orange) when the Yukawa couplings are same order
($y_{\mu\tau}\sim y_{\tau\mu}$).
This is due to the the tree level correction given by
Eq.~(\ref{eq:lep_uni}).
In contrast, when the Yukawa couplings are hierarchical, one of the
loop corrections for the charged lepton and $Z$ boson decays becomes
stronger. 
The gray region surrounded by the dot-dashed line shows the bounds of 
the potential stability. 
In fact, the potential stability bounds are slightly stronger than the
triviality bounds.
This is because we take the negative quartic couplings $\lambda_4$
and $\lambda_5$ at the electroweak scale, and the
Yukawa couplings involved in the $\beta_{\lambda_4}$ and $\beta_{\lambda_5}$ make
$\lambda_4$ and $\lambda_5$ further negative at the cut-off scale if the
Yukawa couplings are $\mathcal{O}(1)$.
As a result, it conflicts with Eq.~(\ref{eq:p_stab}) at the cut-off scale. 
Note that this bound is relaxed if a smaller cut-off scale $\Lambda$ is
assumed.

 \section{Discussions}
\begin{figure}[tb]
\centering 
\includegraphics[width=12cm]{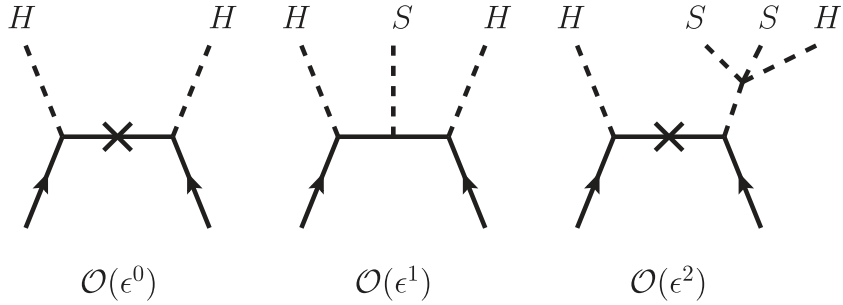}
\caption{Feynman diagrams for neutrino mass generation through seesaw mechanism under the $Z_{4}$ flavor symmetry.}
\label{Fig:SeesawZ4}
\end{figure}

\subsection{Neutrino Mass Generation Sector}
As we mentioned in the beginning, the $Z_{4}$ flavor symmetry in our minimal model 
must be broken in order to fit the observed data of neutrino masses and mixings. 
As a simple example for neutrino mass generation, we here consider the type-I seesaw 
mechanism. 
A SM singlet scalar $S$ with $Z_{4}$ charge $\omega$ and 
a three generation of right-handed neutrinos $(N_{1R}, N_{2R}, N_{3R})$ with 
$(1, \omega, \underline{\omega})$ are introduced to the model.  
The Lagrangian for the neutrino mass generation sector is 
\begin{align}
-{\mathcal L}_{N}
&= 
+\frac12 
\begin{pmatrix} \overline{N_{1R}^{c}} & \overline{N_{2R}^{c}} & \overline{N_{3R}^{c}} \end{pmatrix} 
\begin{pmatrix} M_{1} & y_{12} S^{*} & y_{13}S \\ y_{12} S^{*} & & M_{23} \\ y_{13} S & M_{23} & \end{pmatrix} 
\begin{pmatrix} N_{1R}^{} \\ N_{2R}^{} \\ N_{3R}^{} \end{pmatrix} 
\nonumber \\
& \qquad 
+\begin{pmatrix} \overline{L_{e}} & \overline{L_{\mu}} & \overline{L_{\tau}} \end{pmatrix} 
\begin{pmatrix} y_{e1}^{} \tilde{H} & & \\ & y_{\mu2}^{} \tilde{H} & y_{\mu3} \tilde{\Phi} \\ 
& y_{\tau2} \tilde{\Phi} & y_{\tau3}^{} \tilde{H} \end{pmatrix} 
\begin{pmatrix} N_{1R}^{} \\ N_{2R}^{} \\ N_{3R}^{} \end{pmatrix} 
+\text{H.c.}
\end{align}
%
where $\tilde{X} = i\sigma_{2} X^{*}~(X=H,\Phi)$.
The singlet $S$ is assumed to have a VEV $\epsilon \langle S \rangle$, which breaks 
the $Z_{4}$ symmetry, where $\epsilon$ is introduced to count the order of singlet VEVs. 
At leading order, ${\mathcal O}(\epsilon^{0})$, the (symmetric) neutrino mass matrix has 
non-zero values only in $(1,1)$ and $(2,3)$ elements (see also Fig.~\ref{Fig:SeesawZ4}). 
At this order, due to the vanishing $(2,2)$ and $(3,3)$ elements, 
a large $\theta_{23}$ mixing is naturally obtained in this model. 
At the next leading order, ${\mathcal O}(\epsilon^{1})$, the matrix takes 
the two zero minor structure~\cite{Asai:2017ryy, Asai:2018ocx}. 
This form of the neutrino mass matrix confronts a severe constraint on the sum of 
neutrino masses from cosmological observation~\cite{Aghanim:2018eyx}. 
In our model, a quartic term, $\kappa\, S^{2}H^{\dag}\Phi$, is allowed 
by the $Z_{4}$ flavor symmetry. 
Through this coupling, a small VEV for $\Phi$, i.e., $\langle \Phi \rangle \sim \kappa\, 
\epsilon^{2}(\langle S \rangle^{2}/M_{\phi}^{2})v$ is induced from the singlet VEV. 
As a result, at ${\mathcal O}(\epsilon^{2})$ we have additional contributions to the mass matrix. 
Then, the total structure of the neutrino mass matrix is  
%
\begin{align}
M_{\nu}
\propto 
\begin{pmatrix} {\mathcal O}(\epsilon^{0}) & {\mathcal O}(\epsilon) & {\mathcal O}(\epsilon) \\
{\mathcal O}(\epsilon) & {\mathcal O}(\kappa\,\epsilon^{2}) & {\mathcal O}(\epsilon^{0})\\
{\mathcal O}(\epsilon) & {\mathcal O}(\epsilon^{0}) & {\mathcal O}(\kappa\,\epsilon^{2})
\end{pmatrix}. 
\end{align}
%
Therefore, in the present model, 
the constraints from neutrino data are relatively relaxed as compared 
with the minimal gauged $U(1)_{{\mathbf L}_{\mu}-{\mathbf L}_{\tau}}$ model.

\subsection{Collider Signature}
In the previous section, we have taken into account the direct collider search constraint 
of charged scalars ($m_\phi\gtrsim700~\mathrm{GeV}$). 
This bound will be improved further at the future LHC running by the same search mode. 
In our model, 
the neutral scalars $(\rho, \eta)$ can be lighter than the charged scalar. 
Such light scalars can be produced at the LHC, 
and give interesting distinctive signals. 
Because of the flavor charge conservation in the $Z_{4}$ symmetric limit, 
they are produced in a pair $q\bar{q}\,(e^{+}e^{-})\to \rho\,\eta$ at hadron (lepton) colliders 
and their primary decay modes are $\mu\tau$ pairs.  
So far no dedicated search has been performed, and it was shown 
in the Type-X THDM that $2\mu2\tau$ final states can be approximately reconstructed 
even at the hadron collider~\cite{Kanemura:2011kx}. 
Application of this analysis to our model seems to be easy. 
Firstly, there is no suppression of the signal events by their branching ratio. 
Secondly, thanks to the collinear approximation of tau leptons, 
the LFV invariant mass $M_{\mu\tau}$ is fully reconstructable. 
Then, $M_{\mu\tau}$ is used for a very good discriminant against background events. 
The study for the discovery potential of $(\rho,\eta)$ is 
beyond the scope of this paper, and we leave it for the future.

\subsection{Indirect Signals}
\subsubsection{Muon Electric Dipole Moment}
If the new Yukawa couplings $y_{\mu\tau}$, $y_{\tau\mu}$ are complex,
electric dipole moment (EDM) of muon is induced by the same diagram for 
muon anomalous magnetic moment in Fig.~\ref{fig:2}, which is computed as 
\begin{equation}
 \frac{d_{\mu}}{e}=\frac{\mathrm{Im}\left(y_{\mu\tau}y_{\tau\mu}\right)}{2(4\pi)^2}
  \left[
   \frac{m_\tau}{m_\rho^2}I_2\left(\frac{m_\mu^2}{m_\rho^2},\frac{m_\tau^2}{m_\rho^2}\right)
   -\frac{m_\tau}{m_\eta^2}I_2\left(\frac{m_\mu^2}{m_\eta^2},\frac{m_\tau^2}{m_\eta^2}\right)
	  \right].
  \label{eq:edm}
\end{equation}
Similar to the case of muon magnetic moment, Eq.~(\ref{eq:edm}) has a potentially large 
contribution from the chirality flipping effect. 

The current experimental bound for muon EDM is given by the Muon $g-2$ Collaboration (BNL) as~\cite{Bennett:2008dy}
\begin{align}
 \frac{|d_\mu|}{e}< 1.9\times10^{-19}~\mathrm{cm}.
\end{align}
In addition to the current bound, factor $10$ improvement is expected by the
future FNAL E989 experiment~\cite{Price:2018}, and the future
sensitivity of the J-PARC $g-2$/EDM Collaboration is roughly $|d_\mu|/e\sim10^{-21}~\mathrm{cm}$~\cite{Saito:2012zz}. 

\begin{figure}[t]
\centering
 \includegraphics[scale=0.7]{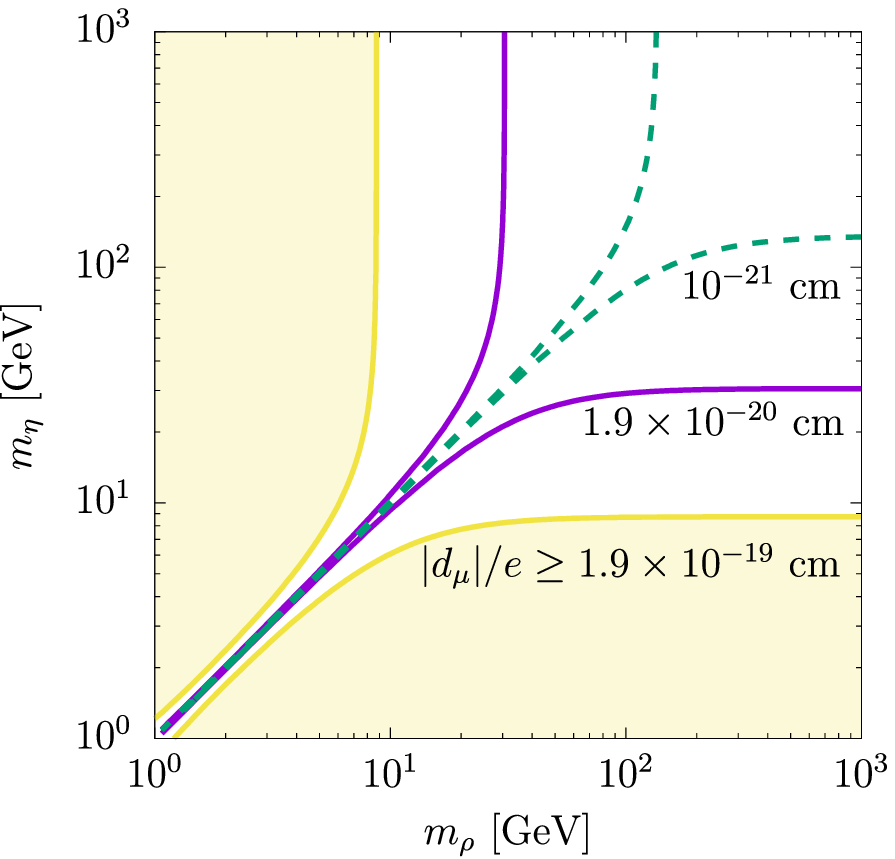}
 \qquad
 \includegraphics[scale=0.7]{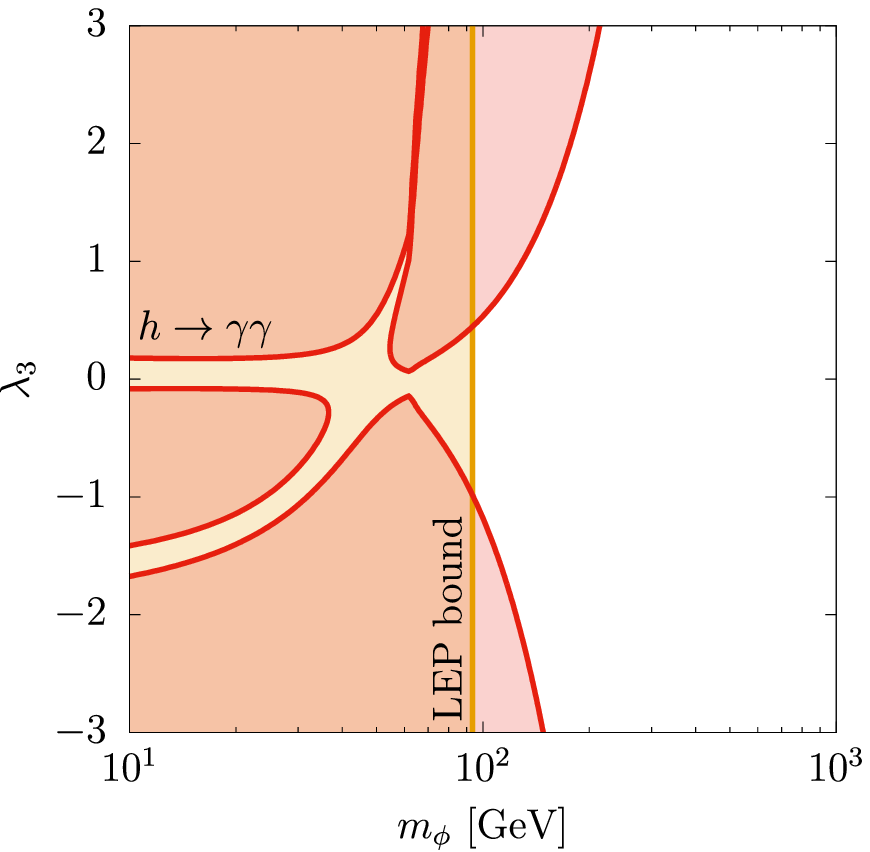}
 \caption{Left: Contours of muon EDM $|d_\mu|/e$ where
 $\mathrm{Im}\left(y_{\mu\tau}y_{\tau\mu}\right)=1$. The yellow
 region is excluded by the Muon $g-2$ Collaboration, and the solid purple
 and dashed green lines are the future prospect of EDM experimental reach. 
 Right: Parameter space excluded by $h\to\gamma\gamma$ signal strength (red)
 and the LEP bound (orange) in ($m_\phi$, $\lambda_3$) plane.}
 \label{fig:edm}
\end{figure}

In the left panel of Fig.~\ref{fig:edm}, 
we give a contour plot of the muon EDM predictions in the ($m_\rho$,
$m_\eta$) plane where we assume $\mathrm{Im}\left(y_{\mu\tau}y_{\tau\mu}\right)=1$. 
The yellow region is already excluded by the current muon EDM limit.
{We see that the current muon EDM limit does not exclude the model 
without requiring the tuning in the imaginary part of the Yukawa couplings.} 
The solid purple and dashed green lines are the future sensitivities of
the FNAL E989 and J-PARC $g-2$/EDM, respectively. 
Although the constraint of the current bound is not so strong, the
future experiments can explore parameter space furthermore.

 \subsubsection{$h\to\gamma\gamma$}
 The additional contribution to $h\to\gamma\gamma$ can appear through the charged scalar loop. 
 The decay amplitude including the SM contribution is computed
 as~\cite{Djouadi:2005gi,Swiezewska:2012eh} 
 \begin{align}
 i\mathcal{M}_{h\to\gamma\gamma}=
  \frac{igm_W\alpha_\mathrm{em}}{(2\pi)\tau_W}\epsilon_{\mu}^*\epsilon_{\nu}^*g^{\mu\nu}
  \left[
   F_1\left(\tau_W\right)
   +\sum_{f}N_cQ_f^2F_{1/2}\left(\tau_f\right)
   +\frac{\lambda_3v^2}{2m_{\phi}^2}F_0\left(\tau_\phi\right)
	  \right],
  \label{eq:hgg}
 \end{align}
 where $\tau_i\equiv 4m_i^2/m_h^2$, $N_c=1,3$ is color factor, $Q_f$ is the electric charge of
 the SM fermions, $\epsilon_{\mu/\nu}$ is the photon polarization vector, 
 and the loop functions $F_{1}(\tau)$,
$F_{1/2}(\tau)$ and $F_0(\tau)$ are given by
\begin{align}
 F_1(\tau)
  &=
  2+3\tau+3\tau\left(2-\tau\right)f(\tau),\\
 F_{1/2}(\tau)
  &=
  -2\tau\Bigl(1+(1-\tau)f(\tau)\Bigr),\\
 F_0(\tau)
  &=
  \tau\Bigl(1-\tau f(\tau)\Bigr),
\end{align}
with
\begin{align}
 f(\tau)=\left\{
	  \begin{array}{cc}
	   \displaystyle\mathrm{arcsin}^2\left(\frac{1}{\sqrt{\tau}}\right)
	    & \text{for}\quad \tau>1\\
	   \displaystyle-\frac{1}{4}\left[\log\left(\frac{1+\sqrt{1-\tau}}{1-\sqrt{1-\tau}}\right)-i\pi\right]^2
	    & \text{for}\quad \tau<1
	  \end{array}
	 \right..
\end{align}
 The last term in Eq.~(\ref{eq:hgg}) corresponds to the
 new contribution which is controlled by the quartic coupling
 $\lambda_3$ in the scalar potential.
Then, the partial decay width is calculated as
\begin{align}
 \Gamma_{h\to\gamma\gamma}=
  \frac{G_{\!F}\alpha_\mathrm{em}^2m_h^3}{128\sqrt{2}\pi^3}
  \left|
      F_1\left(\tau_W\right)
   +\sum_{f}N_cQ_f^2F_{1/2}\left(\tau_f\right)
   +\frac{\lambda_3v^2}{2m_{\phi}^2}F_0\left(\tau_\phi\right)
  \right|^2. 
\end{align}

The signal strength for $h\to\gamma\gamma$ defined by the ratio of the
observed Higgs boson decay to the SM prediction has been reported as
$\mu=0.99_{-0.14}^{+0.15}$ by the
ATLAS Collaboration~\cite{Aaboud:2018xdt}, and $1.18_{-0.14}^{+0.17}$ by the CMS
Collaboration~\cite{Sirunyan:2018ouh}.
The signal strength deviates from unity if non-zero value of the quartic coupling $\lambda_3$ exists. 
The constrained parameter space in the ($m_\phi$, $\lambda_3$) plane is shown 
in the right panel of Fig.~\ref{fig:edm}, where the red region is
excluded by the PDG data $\mu=1.16\pm0.18$ at $2\sigma$
CL~\cite{Tanabashi:2018oca}, and the orange 
region is excluded by the LEP limit $m_{\phi}\lesssim93.5~\mathrm{GeV}$. 
One can see that the parameter space with $|\lambda_3|=\mathcal{O}(1)$ is
ruled out if $100~\mathrm{GeV}\lesssim m_{\phi}\lesssim 200~\mathrm{GeV}$. 
There is no substantial constraint if $m_{\phi}\gtrsim200~\mathrm{GeV}$.

\section{Summary and Conclusions}
\label{sec:6}

We have studied models based on leptonic flavor symmetries, 
which can accommodate the long-standing muon $g-2$ anomaly. 
The minimal model is based on a $Z_4$ lepton flavor symmetry, 
and includes an inert doublet scalar charged under the flavor symmetry. 
Large muon anomalous magnetic moment is realized by the chirality 
enhancement with the factor $m_{\tau}/m_{\mu}\approx17$ in this model. 
We have also analytically formulated the constraints from the electroweak
precision tests and lepton universality. 
Taking into account all these constraints, 
allowed parameter space is explored numerically.
For the electroweak precision tests, it has been found that the
constraint can easily be evaded if the quartic couplings $\lambda_4$ and
$\lambda_5$ are relatively small or the relation
$\lambda_5\sim\pm\lambda_4$ is satisfied, which corresponds to one of
neutral scalars $\rho$ and $\eta$ is nearly degenerate with the charged
scalar $\phi^+$.
For lepton universality, we have computed tree and one-loop corrections 
of heavier charged lepton decays, and one-loop correction for $Z$ boson
decay. 
We have found that the tree level correction becomes dominant when the
Yukawa couplings are comparable ($y_{\mu\tau}\sim y_{\tau\mu}$) while
the loop correction becomes important for hierarchical Yukawa
couplings. 
In addition, we have numerically examined the potential stability
conditions and triviality bounds assuming the cut-off scale of the model,
$\Lambda=100~\mathrm{TeV}$. 
We have successfully found that the parameter region where the discrepancy 
in the muon $g-2$ is explained at $2\sigma$ level while satisfying 
all relevant constraints. 
As further perspective of the minimal $Z_4$ model, neutrino mass
generation with Type-I seesaw mechanism, discriminative collider
signatures, indirect signals from muon EDM and Higgs decay width into
$\gamma\gamma$ have also been discussed. 
We have also found that some parameter space can be explored by the future
EDM experiments if rather large CP phase exists in the Yukawa couplings. 
The signal strength of the Higgs decay width into $\gamma\gamma$ is
influenced by the new contribution if the charged scalar mass is less than $200~\mathrm{GeV}$. 


\section*{Acknowledgments}
\noindent
The work of KT is supported by JSPS Grant-in-Aid for Young Scientists (B) (Grant No. 16K17697), 
by the MEXT Grant-in-Aid for Scientific Research on Innovation Areas (Grant No. 16H00868). 
TT acknowledges funding from the Natural Sciences and Engineering
Research Council of Canada (NSERC).
Numerical computation in this work was carried out at the Yukawa
Institute Computer Facility. 

\section*{Appendix A}
We list the $\beta$ functions for the gauge couplings, quartic
couplings and Yukawa couplings at one loop level, which have been used
to derive the triviality bound.
We have used the public package \texttt{SARAH}~\cite{Staub:2010jh,
Staub:2013tta} to obtain the following analytic expressions. 
Note that the effect of the charged lepton and quark Yukawa couplings
are neglected except for the top Yukawa coupling. 
\\

\noindent\underline{\bf $\beta$ functions for gauge couplings:} 
\begin{align}
 \beta_{g^{\prime}}&= 7{g^\prime}^3,\\
 \beta_{g}&= -3g^3,\\
 \beta_{g_c^{}}&= -7g_c^3.
\end{align}

\noindent\underline{\bf $\beta$ functions for quartic couplings:} 
\begin{align}
 \beta_{\lambda_{1}}&=
 \frac{3}{8}{g^\prime}^4+\frac{3}{4}{g^\prime}^2g^2+\frac{9}{8}g^4
 -3\lambda_1\left({g^\prime}^2+3g^2\right)+24\lambda_1^2+2\lambda_3^2+2\lambda_3\lambda_4+\lambda_4^2+\lambda_5^2\nonumber\\
 &~~~+12\lambda_1y_t^2
 -6y_t^4,\\
 \beta_{\lambda_2}&=
 \frac{3}{8}{g^\prime}^4+\frac{3}{4}{g^\prime}^2g^2+\frac{9}{8}g^4
 -3\lambda_2\left({g^\prime}^2+3g^2\right)+24\lambda_2^2+2\lambda_3^2+2\lambda_3\lambda_4+\lambda_4^2+\lambda_5^2\nonumber\\
 &~~~+4\lambda_2\left(|y_{\mu\tau}|^2+|y_{\tau\mu}|^2\right)
 -2\left(|y_{\mu\tau}|^4+|y_{\tau\mu}|^4\right),\\
 \beta_{\lambda_{3}}&=
 \frac{3}{4}{g^\prime}^4+\frac{3}{2}{g^\prime}^2g^2+\frac{9}{4}g^4
 -3\lambda_3\left({g^\prime}^2+3g^2\right)+4\left(\lambda_1+\lambda_2\right)\left(3\lambda_3+\lambda_4\right)\nonumber\\
 &~~~+4\lambda_3^2+2\lambda_4^2+10\lambda_5^2
 +2\lambda_3\left(|y_{\mu\tau}|^2+|y_{\tau\mu}|^2
 +3y_t^2\right),\\
 \beta_{\lambda_{4}}&=
 -3{g^\prime}^2g^2-3\lambda_4\left({g^\prime}^2+3g^2\right)
 +4\left(\lambda_1+\lambda_2+2\lambda_3+\lambda_4\right)\lambda_4-8\lambda_5^2\nonumber\\
 &~~~+2\lambda_4\left(|y_{\mu\tau}|^2+|y_{\tau\mu}|^2+3y_t^2\right),\\
 \beta_{\lambda_{5}}&=
 -3\left({g^\prime}^2+3g^2\right)\lambda_5+4\left(\lambda_1+\lambda_2+2\lambda_3-\lambda_4\right)\lambda_5
 +2\lambda_5\left(|y_{\mu\tau}|^2+|y_{\tau\mu}|^2
 +3y_t^2\right).
\end{align}

\noindent\underline{\bf $\beta$ functions for Yukawa couplings:}
\begin{align}
 \beta_{y_{\mu\tau}}&=
 \left[\frac{5}{2}|y_{\mu\tau}|^2+|y_{\tau\mu}|^2
 -\frac{9}{4}\left(\frac{5}{3}{g^\prime}^2+g^2\right)
 \right]y_{\mu\tau},\\
 \beta_{y_{\tau\mu}}&=
 \left[|y_{\mu\tau}|^2+\frac{5}{2}|y_{\tau\mu}|^2
 -\frac{9}{4}\left(\frac{5}{3}{g^\prime}^2+g^2\right)
 \right]y_{\tau\mu},\\
 \beta_{y_t}&=
 \left[-\frac{17}{12}{g^\prime}^2-\frac{9}{4}g^2-8g_c^2
 +\frac{9}{2}y_t^2\right]y_t.
\end{align}
\\



\begin{thebibliography}{200}

\bibitem{Tanabashi:2018oca} 
  M.~Tanabashi {\it et al.} [Particle Data Group],
  Phys.\ Rev.\ D {\bf 98}, no. 3, 030001 (2018).


\bibitem{Keshavarzi:2018mgv} 
  A.~Keshavarzi, D.~Nomura and T.~Teubner,
  Phys.\ Rev.\ D {\bf 97}, no. 11, 114025 (2018)
  [arXiv:1802.02995 [hep-ph]].
  

\bibitem{Prades:2009tw} 
  J.~Prades, E.~de Rafael and A.~Vainshtein,
  Adv.\ Ser.\ Direct.\ High Energy Phys.\  {\bf 20}, 303 (2009)
  [arXiv:0901.0306 [hep-ph]].


\bibitem{Chapelain:2017syu} 
  A.~Chapelain [Muon g-2 Collaboration],
  EPJ Web Conf.\  {\bf 137}, 08001 (2017)
  [arXiv:1701.02807 [physics.ins-det]].



\bibitem{Abe:2017jqo} 
  T.~Abe, R.~Sato and K.~Yagyu,
  JHEP {\bf 1707}, 012 (2017)
  [arXiv:1705.01469 [hep-ph]].

\bibitem{Chun:2016hzs} 
  E.~J.~Chun and J.~Kim,
  JHEP {\bf 1607}, 110 (2016)
  [arXiv:1605.06298 [hep-ph]].

\bibitem{Abe:2015oca} 
  T.~Abe, R.~Sato and K.~Yagyu,
  JHEP {\bf 1507}, 064 (2015)
  [arXiv:1504.07059 [hep-ph]].


\bibitem{Crivellin:2019dun} 
  A.~Crivellin, D.~M\"uller and C.~Wiegand,
  arXiv:1903.10440 [hep-ph].


\bibitem{Omura:2015nja} 
  Y.~Omura, E.~Senaha and K.~Tobe,
  JHEP {\bf 1505}, 028 (2015)
  [arXiv:1502.07824 [hep-ph]].

\bibitem{Omura:2015xcg} 
  Y.~Omura, E.~Senaha and K.~Tobe,
  Phys.\ Rev.\ D {\bf 94}, no. 5, 055019 (2016)
  [arXiv:1511.08880 [hep-ph]].


\bibitem{Baek:2001kca} 
  S.~Baek, N.~G.~Deshpande, X.~G.~He and P.~Ko,
  Phys.\ Rev.\ D {\bf 64}, 055006 (2001)
  [hep-ph/0104141].

\bibitem{Ma:2001md} 
  E.~Ma, D.~P.~Roy and S.~Roy,
  Phys.\ Lett.\ B {\bf 525}, 101 (2002)
  [hep-ph/0110146].

\bibitem{Endo:2012hp} 
  M.~Endo, K.~Hamaguchi and G.~Mishima,
  Phys.\ Rev.\ D {\bf 86}, 095029 (2012)
  [arXiv:1209.2558 [hep-ph]].

\bibitem{Endo:2013bba} 
  M.~Endo, K.~Hamaguchi, S.~Iwamoto and T.~Yoshinaga,
  JHEP {\bf 1401}, 123 (2014)
  [arXiv:1303.4256 [hep-ph]].

\bibitem{Marciano:2016yhf} 
  W.~J.~Marciano, A.~Masiero, P.~Paradisi and M.~Passera,
  Phys.\ Rev.\ D {\bf 94}, no. 11, 115033 (2016)
  [arXiv:1607.01022 [hep-ph]].

\bibitem{BarShalom:2011bb} 
  S.~Bar-Shalom, S.~Nandi and A.~Soni,
  Phys.\ Lett.\ B {\bf 709}, 207 (2012)
  [arXiv:1112.3661 [hep-ph]].


\bibitem{Chiang:2017vcl} 
  C.~W.~Chiang and K.~Tsumura,
  JHEP {\bf 1805}, 069 (2018)
  [arXiv:1712.00574 [hep-ph]].


\bibitem{Asai:2017ryy} 
  K.~Asai, K.~Hamaguchi and N.~Nagata,
  Eur.\ Phys.\ J.\ C {\bf 77}, no. 11, 763 (2017)
  [arXiv:1705.00419 [hep-ph]].


\bibitem{Deshpande:1977rw} 
  N.~G.~Deshpande and E.~Ma,
  Phys.\ Rev.\ D {\bf 18}, 2574 (1978).

\bibitem{Hambye:2009pw} 
  T.~Hambye, F.-S.~Ling, L.~Lopez Honorez and J.~Rocher,
  JHEP {\bf 0907}, 090 (2009)
  Erratum: [JHEP {\bf 1005}, 066 (2010)]
  [arXiv:0903.4010 [hep-ph]].





\bibitem{Peskin:1991sw} 
  M.~E.~Peskin and T.~Takeuchi,
  Phys.\ Rev.\ D {\bf 46}, 381 (1992).
  
  
\bibitem{Barbieri:2006dq} 
  R.~Barbieri, L.~J.~Hall and V.~S.~Rychkov,
  Phys.\ Rev.\ D {\bf 74}, 015007 (2006)
  [hep-ph/0603188].


\bibitem{Peskin:2001rw} 
  M.~E.~Peskin and J.~D.~Wells,
  Phys.\ Rev.\ D {\bf 64}, 093003 (2001)
  [hep-ph/0101342].
  
  
\bibitem{Kanemura:2011sj} 
  S.~Kanemura, Y.~Okada, H.~Taniguchi and K.~Tsumura,
  Phys.\ Lett.\ B {\bf 704}, 303 (2011)
  [arXiv:1108.3297 [hep-ph]].


\bibitem{Baak:2014ora} 
  M.~Baak {\it et al.} [Gfitter Group],
  Eur.\ Phys.\ J.\ C {\bf 74}, 3046 (2014)
  [arXiv:1407.3792 [hep-ph]].



\bibitem{Marciano:1988vm} 
  W.~J.~Marciano and A.~Sirlin,
  Phys.\ Rev.\ Lett.\  {\bf 61}, 1815 (1988).




\bibitem{Amhis:2014hma} 
  Y.~Amhis {\it et al.} [Heavy Flavor Averaging Group (HFAG)],
  arXiv:1412.7515 [hep-ex].



\bibitem{ALEPH:2005ab} 
  S.~Schael {\it et al.} [ALEPH and DELPHI and L3 and OPAL and SLD
	Collaborations and LEP Electroweak Working Group and SLD
	Electroweak Group and SLD Heavy Flavour Group], 
  Phys.\ Rept.\  {\bf 427}, 257 (2006)
  [hep-ex/0509008].


\bibitem{Abbiendi:2013hk} 
  G.~Abbiendi {\it et al.} [ALEPH and DELPHI and L3 and OPAL and LEP Collaborations],
  Eur.\ Phys.\ J.\ C {\bf 73}, 2463 (2013)
  [arXiv:1301.6065 [hep-ex]].

\bibitem{Aaboud:2018jiw} 
  M.~Aaboud {\it et al.} [ATLAS Collaboration],
  Eur.\ Phys.\ J.\ C {\bf 78}, no. 12, 995 (2018)
  [arXiv:1803.02762 [hep-ex]].

\bibitem{ATLAS:2019cfv} 
  The ATLAS collaboration [ATLAS Collaboration],
  ATLAS-CONF-2019-008.

\bibitem{Asai:2018ocx} 
  K.~Asai, K.~Hamaguchi, N.~Nagata, S.~Y.~Tseng and K.~Tsumura,
  Phys.\ Rev.\ D {\bf 99}, no. 5, 055029 (2019)
  [arXiv:1811.07571 [hep-ph]].


\bibitem{Aghanim:2018eyx} 
  N.~Aghanim {\it et al.} [Planck Collaboration],
  arXiv:1807.06209 [astro-ph.CO].


\bibitem{Kanemura:2011kx} 
  S.~Kanemura, K.~Tsumura and H.~Yokoya,
  Phys.\ Rev.\ D {\bf 85}, 095001 (2012)
  [arXiv:1111.6089 [hep-ph]].


\bibitem{Bennett:2008dy} 
  G.~W.~Bennett {\it et al.} [Muon (g-2) Collaboration],
  Phys.\ Rev.\ D {\bf 80}, 052008 (2009)
  [arXiv:0811.1207 [hep-ex]].


 \bibitem{Price:2018}
 J.~Price, talk at ``Workshop on future muon EDM searches at Fermilab
	 and worldwide'', \url{https://indico.fnal.gov/event/18239/}.


\bibitem{Saito:2012zz} 
  N.~Saito [J-PARC g-'2/EDM Collaboration],
  AIP Conf.\ Proc.\  {\bf 1467}, 45 (2012).



\bibitem{Djouadi:2005gi} 
  A.~Djouadi,
  Phys.\ Rept.\  {\bf 457}, 1 (2008)
  [hep-ph/0503172].

\bibitem{Swiezewska:2012eh} 
  B.~Swiezewska and M.~Krawczyk,
  Phys.\ Rev.\ D {\bf 88}, no. 3, 035019 (2013)
  [arXiv:1212.4100 [hep-ph]].


\bibitem{Aaboud:2018xdt} 
  M.~Aaboud {\it et al.} [ATLAS Collaboration],
  Phys.\ Rev.\ D {\bf 98}, 052005 (2018)
  [arXiv:1802.04146 [hep-ex]].


\bibitem{Sirunyan:2018ouh} 
  A.~M.~Sirunyan {\it et al.} [CMS Collaboration],
  JHEP {\bf 1811}, 185 (2018)
  [arXiv:1804.02716 [hep-ex]].



\bibitem{Staub:2010jh} 
  F.~Staub,
  Comput.\ Phys.\ Commun.\  {\bf 182}, 808 (2011)
  [arXiv:1002.0840 [hep-ph]].

\bibitem{Staub:2013tta} 
  F.~Staub,
  Comput.\ Phys.\ Commun.\  {\bf 185}, 1773 (2014)
  [arXiv:1309.7223 [hep-ph]].
\end{thebibliography}
\end{document}